\documentclass[a4paper,twoside]{article}

\usepackage{paralist}
\usepackage{cite}
\usepackage{comment}
\usepackage{epsfig}
\usepackage{subcaption}
\usepackage{calc}
\usepackage{amssymb}
\usepackage{amstext}
\usepackage{amsmath}
\usepackage{amsthm}
\usepackage{multicol}
\usepackage{pslatex}
\usepackage{apalike}
\usepackage[bottom]{footmisc}
\usepackage{url}
\usepackage{graphics}
\usepackage{listings}
\usepackage{xcolor}
\usepackage{tikz}
\usepackage{hyperref}
\usepackage{SCITEPRESS}     % Please add other packages that you may need BEFORE the SCITEPRESS.sty package.

\DeclareFixedFont{\ttb}{T1}{txtt}{bx}{n}{6.5} % for bold
\DeclareFixedFont{\ttm}{T1}{txtt}{m}{n}{6.5}  % for normal

\definecolor{deepblue}{rgb}{0,0,0.5}
\definecolor{deepred}{rgb}{0.6,0,0}
\definecolor{deepgreen}{rgb}{0,0.5,0}

\usetikzlibrary{arrows,trees,backgrounds,automata,shapes,decorations,plotmarks,fit,calc,positioning,shadows,chains}
\tikzstyle{every pin edge}=[<-,shorten <=1pt]
\tikzstyle{neuron}=[circle,fill=black!25,minimum size=17pt,inner sep=0pt]
\tikzstyle{input neuron}=[neuron, fill=green!50]
\tikzstyle{output neuron}=[neuron, fill=red!50]
\tikzstyle{hidden neuron}=[neuron, fill=blue!50]
\tikzstyle{annot} = [text width=6em, text centered]
\usetikzlibrary{calc}

\usepackage{xspace}
\newcommand{\relu}{\text{ReLU}\xspace{}}
\newcommand{\true}{\text{True}\xspace{}}
\newcommand{\lts}{\text{LTS}\xspace{}}
\newcommand{\modifier}{\text{modifier}\xspace{}}

\renewcommand\j[1]{\textsc{#1}}

\newcommand{\ra}{\rangle}

\begin{document}

\title{Enhancing Deep Learning with Scenario-Based\\Override Rules: a Case Study}

		\author{\authorname{Adiel Ashrov\sup{1}\orcidAuthor{0000-0003-4510-5335}, Guy Katz\sup{1}\orcidAuthor{0000-0001-5292-801X}}
		\affiliation{\sup{1}The Hebrew University of Jerusalem, Jerusalem, Israel}
		\email{adiel.ashrov@mail.huji.ac.il, guykatz@cs.huji.ac.il}}

	
	\abstract{Deep neural networks (DNNs) have become a crucial
          instrument in the software development toolkit, due to their
          ability to efficiently solve complex problems. Nevertheless,
          DNNs are highly opaque, and can behave in an
          unexpected manner when they encounter unfamiliar input. One
          promising approach for addressing this challenge is by
          extending DNN-based systems with hand-crafted \emph{override
            rules}, which override the DNN's output when certain
          conditions are met. Here, we advocate crafting such override
          rules using the well-studied \emph{scenario-based modeling}
          paradigm, which produces rules that are simple, extensible,
          and powerful enough to ensure the safety of the DNN, while
          also rendering the system more translucent. We report on two
          extensive case studies, which demonstrate the feasibility of
          the approach; and through them, propose an extension to
          scenario-based modeling, which facilitates its integration
          with DNN components. We regard this work as a step towards
          creating safer and more reliable DNN-based systems and models.}

	\begin{comment}
	  \abstract{The abstract should summarize the contents of the paper and should contain at least 70 and at most 200 words. The text must be set to 9-point font size.}
	\end{comment}

	\onecolumn \maketitle \normalsize \setcounter{footnote}{0} \vfill

\section{\uppercase{Introduction}}
\label{sec:introduction}

% DNNs are revolutionizing the world
\emph{Deep learning (DL)} algorithms have been revolutionizing the
world of Computer Science, by enabling engineers to automatically
generate software systems that achieve excellent
performance~\cite{GoBeCo16}. DL algorithms can generalize
examples of the desired behavior of a system into an artifact called a
\emph{deep neural network (DNN)}, whose performance often exceeds that
of manually
designed software~\cite{SiZi14,SiHuMaGuSiVaScAnPaLaDi16}. DNNs are
now being extensively used in domains such as game playing~\cite{MnKaSiGrAnWiRi13},
natural language processing~\cite{CoWeBoKaKaKu11}, protein folding~\cite{JuEvPrGrFiRoTuBaZiPo21}, and many
others.  In addition, they are also being used as controllers within
\emph{critical reactive systems}, such as autonomous cars and unmanned
aircraft~\cite{BoDeDwFiFlGoJaMoMuZh16,JuLoBrOwKo16}.

% The drawbacks of DNNs and DRLs
Although systems powered by DNNs have achieved unprecedented results, there is 
still room for improvement. DNNs are trained automatically, and are highly 
\emph{opaque} --- meaning that we do not comprehend, and cannot reason about, 
their decision-making process. This inability is a cause for concern, as DNNs 
do not always generalize well, and can make severe mistakes. For example, it 
has been observed that state-of-the-art DNNs for traffic sign recognition can
misclassify ``stop'' signs, even though they have been trained on
millions of street images~\cite{PaMcGoJhCeSw17}. When DNNs are deployed in 
reactive systems that are \emph{safety critical}, such mistakes could 
potentially endanger human lives. It is therefore necessary to enhance the 
safety and dependability of these systems, prior to their deployment in the 
field.

% Previous work - guard using scenario based modeling
One appealing approach for bridging the gap between the high
performance of DNNs and the required level of reliability is to
\emph{guard} DNNs with additional, hand-crafted components, which
could override the DNNs in case of clear
mistakes~\cite{ShShSh16,AvBlChHeKoPr19}. This, in turn, raises the
question of how to design and implement these guard components. More
recent work~\cite{HaMaSi22,Ka20b,Ka20a} has suggested fusing DL with modern 
\emph{software engineering (SE)} paradigms, in a
way that would allow for improving the development process, user
experience, and overall safety of the resulting systems. The idea is
to enable domain experts to efficiently and conveniently pour their
knowledge into the system, in the form of hand-crafted modules that
will guarantee that unexpected behavior is avoided.

Here, we focus on one particular mechanism for producing such guard
rules, through the \emph{scenario-based modeling (SBM)}
paradigm~\cite{HaMaWe12}. SBM is a software development paradigm,
whose goal is to enable users to model systems in a way that is
aligned with how they are perceived by humans~\cite{GoMaMe12}. In SBM, the user 
specifies \emph{scenarios}, each of which represents a single desirable or 
undesirable system behavior. These scenarios are fully executable, and can be 
interleaved together at runtime in order to produce cohesive system behavior.
Various studies have shown that SBM is particularly suited for modeling reactive
systems~\cite{BaWeSh18}; and in
particular, reactive systems that involve DNN
components~\cite{YeAmElHaKaMa22,CoYeAmFaHaKa22}.

% Paper goal and Proof of concept applications - Aurora + Robotis
The research questions that we tackled in this work are:
\begin{enumerate}
	\item Can the approach of integrating SBM and DL be applied to  
	state-of-the-art deep learning projects?
	\item Are there idioms that, if added to SBM, could facilitate this 
	integration?
\end{enumerate}
To answer these questions, we apply SBM to guard
two reactive systems powered by deep learning: (1)
\emph{Aurora}~\cite{JaRoGoScTa19}, a \emph{congestion control} protocol whose
goal is to optimize the communication throughput of a computer
network; and (2) the Turtlebot3
platform~\cite{NaShVa21}, a mobile robot capable of performing
\emph{mapless navigation}  towards a predefined target through the use of a
pre-trained DNN as its policy. In both cases, we instrument the DNN core of the 
system with an SBM harness; and then introduce guard scenarios for overriding 
the DNN's outputs in various undesirable situations. In both case studies, we 
demonstrate that our SBM components can indeed enforce various safety goals.
The answer to our first research question is therefore positive, since these
initial results demonstrate the applicability and usefulness of this approach.

% Challanges encoutered
As part of our work on the Aurora and Turtlebot3 systems, we observed
that the integration between the underlying DNNs and SBM components
was not always straightforward. One recurring challenge, which the SBM
framework could only partially tackle, was the need for the SB model
to react immediately, in the same time step, to the decisions made by
the DNN --- as opposed to only reacting to actions that occurred in
previous time steps~\cite{HaMaWe12}. This issue could be
circumvented, but this entailed using ad hoc solutions that go against
the grain of SBM. This observation answers our second research question: indeed,
certain enhancements to SBM are necessary to facilitate a more seamless 
combination of SBM and DL. In order to overcome this difficulty in a more
principled way, we propose here an extension to the SBM framework with
a new type of scenario, which we refer to as a \emph{modifier}
scenario. This extension enabled us to create a cleaner and more
maintainable scenario-based model to guard the DNNs in question. We
describe the experience of using the new kind of scenario, and provide
a formal extension to SBM that includes it.

% Paper structure
The rest of the paper is organized as follows. In
Sec.~\ref{sec:background} we provide the necessary background on DNNs,
and their guarding using SBM. In Sec.~\ref{sec:case_aurora} and
Sec.~\ref{sec:case_robotis} we describe our two case studies.  Next,
in Sec.~\ref{sec:modifier-scenario} we present our extension to SBM,
which supports modifier scenarios.  We follow with a discussion of
related work in Sec.~\ref{sec:related-work}, and conclude in
Sec.~\ref{sec:conclusion}.
	
\section{\uppercase{Background}}
\label{sec:background}
\noindent
\subsection{Deep Reinforcement Learning}
\label{sec:background:dnns}

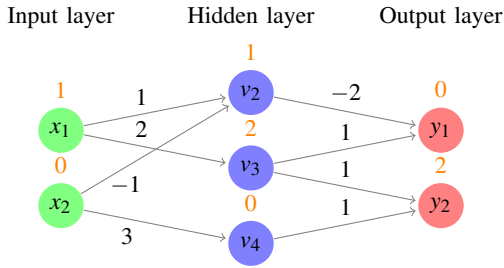
\begin{figure}[htb]
  \begin{center}
    \scalebox{1} {
      \def\layersep{2.5cm}
      \begin{tikzpicture}[shorten >=1pt,->,draw=black!50, node distance=\layersep,font=\footnotesize]
        
        \node[input neuron, label=above:\textcolor{orange}{1}] (I-1) at (0,-1cm) {$x_1$};
        \node[input neuron, label=above:\textcolor{orange}{0}] (I-2) at (0,-2cm) {$x_2$};
        
        \path[yshift=0.5cm] node[hidden neuron, label=above:\textcolor{orange}{1}] (H-1)
        at (\layersep,-1cm) {$v_2$};
        \path[yshift=0.5cm] node[hidden neuron, label=above:\textcolor{orange}{2}] (H-2)
        at (\layersep,-2cm) {$v_3$};
        \path[yshift=0.5cm] node[hidden neuron, label=above:\textcolor{orange}{0}] (H-3)
        at (\layersep,-3cm) {$v_4$};
        
        \node[output neuron, label=above:\textcolor{orange}{0}] at (2*\layersep, -1) (O-1) {$y_1$};
        \node[output neuron, label=above:\textcolor{orange}{2}] at (2*\layersep, -2) (O-2) {$y_2$};
        
        % Connect every node in the hidden layer with the output layer
        \path (I-1) edge[] node[above,pos=0.4] {$1$} (H-1);
        \path (I-1) edge[] node[above,pos=0.4] {$2$} (H-2);
        \path (I-2) edge[] node[below,pos=0.3] {$-1$} (H-1);
        \path (I-2) edge[] node[below,pos=0.3] {$3$} (H-3);
        
        \path (H-1) edge[] node[above] {$-2$} (O-1);
        \path (H-2) edge[] node[above] {$1$} (O-1);
        \path (H-2) edge[] node[above] {$1$} (O-2);
        \path (H-3) edge[] node[above] {$1$} (O-2);
        
        % Annotate the layers
        \node[annot,above of=H-1, node distance=1cm] (hl) {Hidden layer};
        \node[annot,left of=hl] {Input layer};
        \node[annot,right of=hl] {Output layer};
      \end{tikzpicture}
    }
%    \captionsetup{size=small}
    \caption{A small neural network. In orange: the values computed
      for each neuron, for input $(1,0)$.}
    \label{fig:dnn_running_example}
  \end{center}
\end{figure}

% Description of DNNs and DRLs
At a high level, a neural network $N$ can be regarded as a transformation
that maps an input vector $x$ into an output vector $N(x)$. For example, the
small network 
depicted in Fig.~\ref{fig:dnn_running_example} has an
input layer, a single hidden layer, and an output layer. After the input
nodes are assigned values by the user, the assignment of each
consecutive layer's nodes is computed iteratively, as a weighted sum
of neurons from its preceding layer, followed by an activation
function. For the network in Fig.~\ref{fig:dnn_running_example}, the
activation function in use is $y=\relu(x)=\max(0,x)$~\cite{NaHi10}. For 
example, for an input vector $x=(x_1, x_2)$, and an assignment $x_1=1, x_2=0$, 
this process results in the output neurons being assigned the values 
$y_1=0,y_2=2$, and $N(x)=(y_1, y_2)$. If the network 
acts as a classifier, we slightly abuse notation and associate each label with 
the corresponding output neuron. In this
case, since $y_2$ has the higher score, input $x=(1,0)$ is classified as
the label $y_2$. For additional background on DNNs, 
see, e.g.,~\cite{GoBeCo16}.

% DRL - high level description
One method for producing DNNs is via \emph{deep reinforcement learning
  (DRL)}~\cite{SuBa18}. In DRL, an agent is trained
to interact with an environment. Each time, the agent selects an
action, with the goal of maximizing a predetermined reward
function. The process can be regarded as a Markov decision process
(MDP), where the agent attempts to learn a policy for maximizing its returns.
DRL algorithms are used to train DNNs to learn optimal policies, through trial 
and error. DRL has shown excellent results in the context of video games, 
robotics, and in various safety-critical systems such as autonomous driving and
flight control~\cite{SuBa18}.

\begin{figure}[htp]
  \centering
  \includegraphics[width=0.7\linewidth]{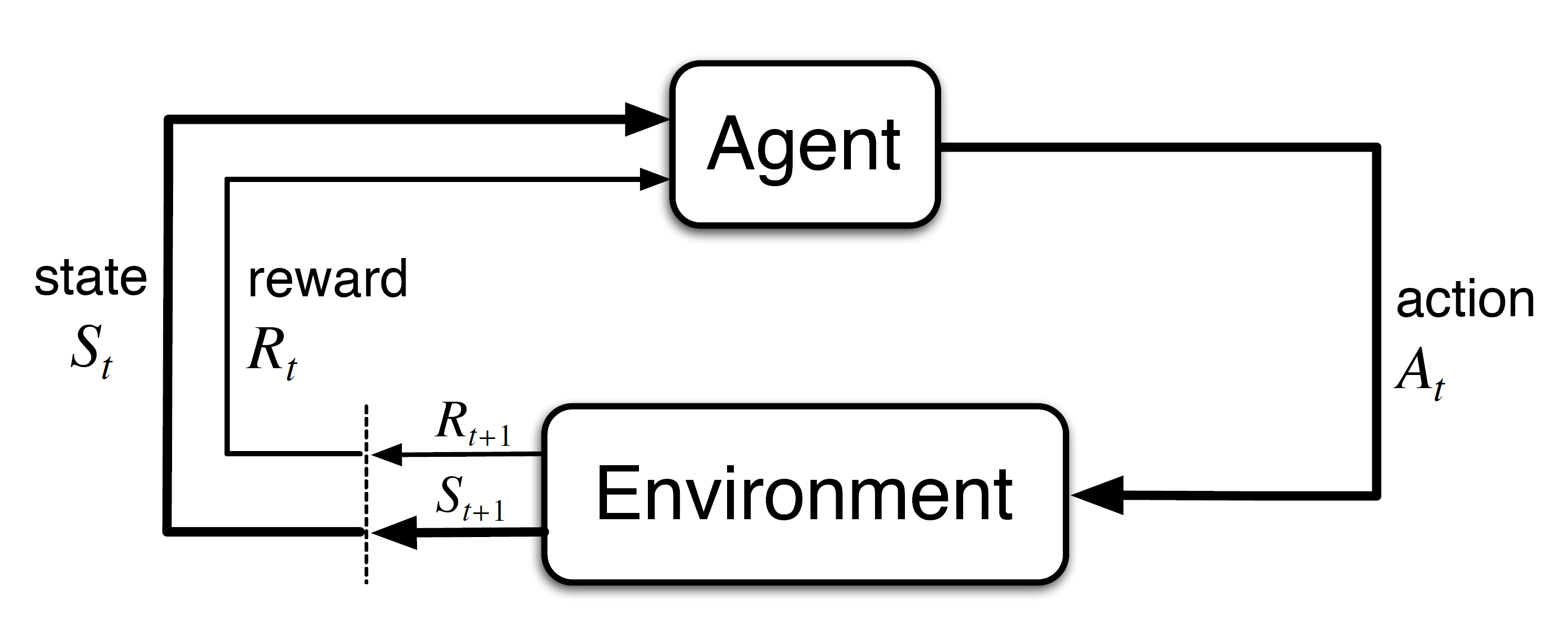}
  \caption{The agent-environment interaction in reinforcement learning
    (borrowed from~\cite{SuBa18}).}
  \label{fig:agent-environment-interaction}
\end{figure}

% DRL agent interaction and training
Fig.~\ref{fig:agent-environment-interaction} describes the basic
interaction between a DRL agent and its environment. At time step $t$,
the agent examines the environment's state $s_{t}$, and chooses an
action $a_{t}$ according to its current policy. At time step ${t+1}$,
and following the selected action $a_{t}$, the agent receives a reward
$R_{t}=R(s_{t}, a_{t})$. The environment then shifts to state
$s_{t+1}$ where the process is repeated. A DRL algorithm trains a DNN
to learn an optimal policy for this
interaction.
	
\subsection{Override Rules}
\label{sec:background-overrideRules}

% Override rules
Given a DNN $N$, an \emph{override rule}~\cite{Ka20a} is defined as a triple $\langle P, Q, \alpha\rangle$, where:
\begin{itemize}
	\item $P$ is a predicate over the network's input $x$.
	\item $Q$ is a predicate over the network's output $N(x)$.
	\item $\alpha$ is an override action.
\end{itemize}
The semantics of an override rule is that if $P(x)$ and $Q(N(x))$ evaluate to 
$\true{}$ for the current input $x$ and the network calculation $N(x)$, then 
the output action $\alpha$ should be selected ---
notwithstanding of the network's output. For example, for the
network from Fig.~\ref{fig:dnn_running_example}, we might define the
following rule:
\[
  \langle x_1 > x_2, \true, y_1 \rangle
\]
We previously saw that for inputs $x_1=1, x_2=0$, the network selects
the label corresponding to $y_2$.  However, if we enforce this
override rule, the selection will be modified to $y_1$. This is
because this particular input satisfies the rule's
conditions (note that $Q=\true$ means that there are no restrictions
on the DNN's output). By adjusting $P$ and $Q$, this
formulation can express a large variety of rules~\cite{Ka20a}.

\subsection{Scenario-Based Modeling}

% General description of SBM
Scenario-based modeling (SBM)~\cite{HaMaWe12}, also known as
\emph{behavioral programming} (\emph{BP}), is a paradigm for modeling
complex reactive systems. The approach is focused on enabling users to
naturally model their perception of the system's
requirements~\cite{GoMaMe12}. At the center of this approach lies the
concept of a \emph{scenario object}: a depiction of a single behavior,
either desirable or undesirable, of the system being modeled. Each
scenario object is created separately, and has no direct contact with
the other scenarios. Rather, it communicates with a global execution
mechanism, which can execute a set of scenarios in a manner that
produces cohesive global behavior.

% Specific description of the execution 
More specifically, a scenario object can be viewed as a transition
system, whose states are referred to as \emph{synchronization
  points}. When the scenario reaches a synchronization point, it
suspends and declares which events it would like to trigger
(\emph{requested events}), which events are forbidden from its
perspective (\emph{blocked events}), and which events it does not
explicitly request, but would like to be notified should they be
triggered (\emph{waited-for events}). The execution infrastructure
waits for all the scenarios to synchronize (or for a subset
thereof~\cite{HaKaKa13}), and selects an event that is requested and
not blocked for triggering. The mechanism then notifies the scenarios
requesting/waiting-for this event that it has been triggered. The
notified scenarios proceed with their execution until reaching the
next synchronization point, where the process is repeated.

\begin{figure*}[htp]
  \scalebox{0.7}{
    \begin{subfigure}{.4\textwidth}
      \includegraphics[scale=0.55]{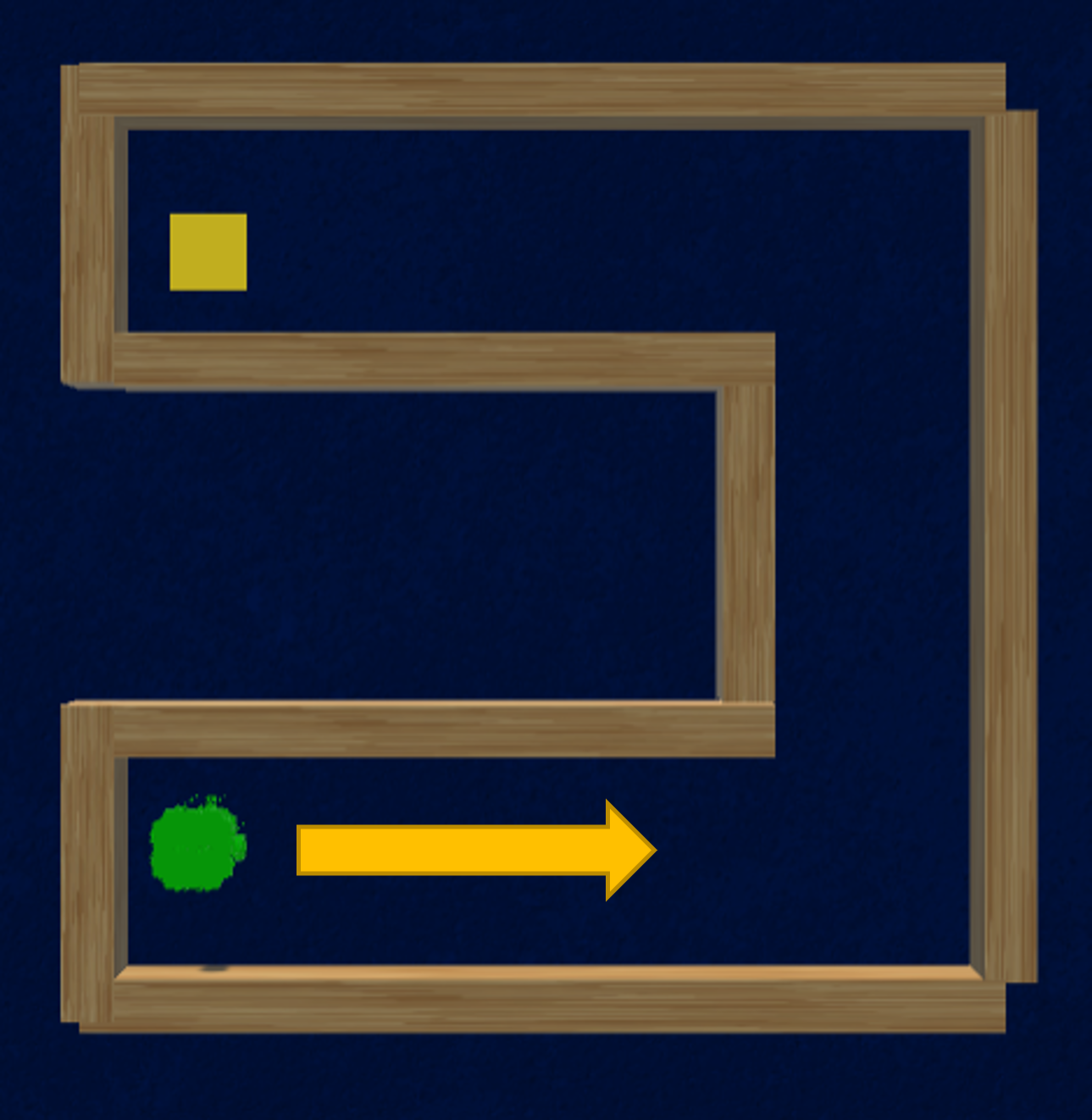}
    \end{subfigure}%
    \begin{subfigure}{.4\textwidth}
      \includegraphics[scale=0.55]{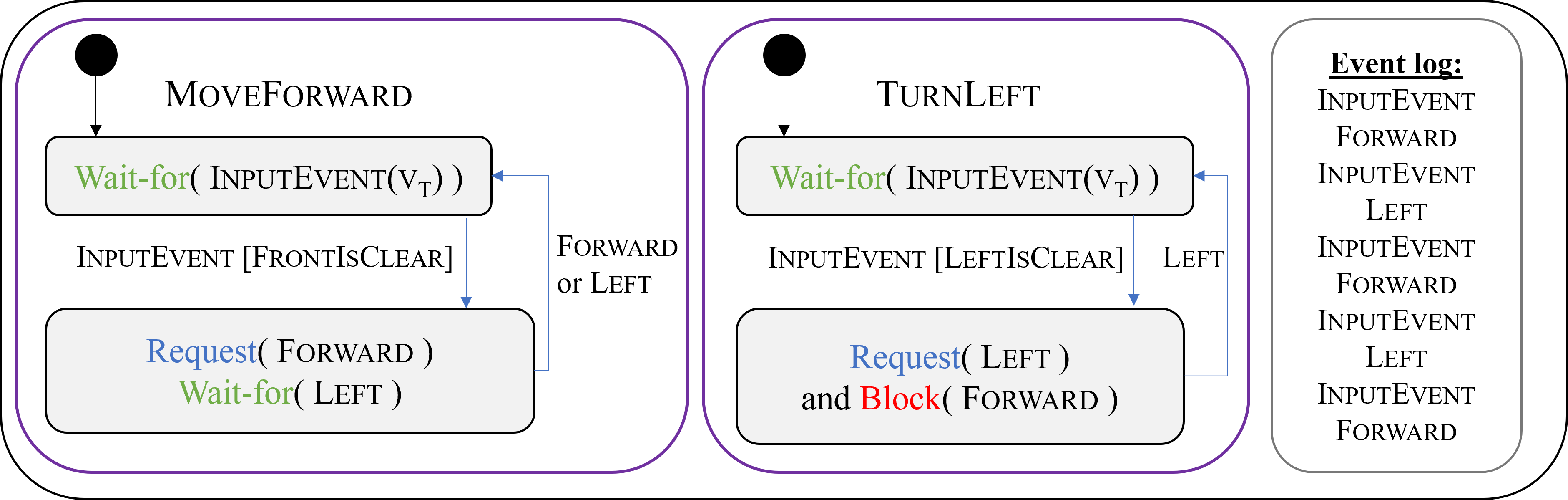}
    \end{subfigure}
  }
  \caption{On the left, a screenshot of the Turtlebot simulator, where
    the robot is headed right and the target appears in the top left
    corner. In the middle, the scenario-based model, written in
    Statechart-like transition systems~\cite{Ha86} extended with 
    SBM. The model contains two scenarios: The \j{MoveForward} scenario and the
    \j{TurnLeft} scenario. The black circles specify the initial
    state. In each state the scenario can request, wait-for or block
    events. Once a requested/waited-for event is triggered, the
    scenario transitions to the appropriate state (highlighted by a
    connecting edge with the event name and an optional Boolean
    condition). On the right, a log of the triggered events during the
    execution, for this particular maze.}
  \label{fig:sbm-maze-example}
\end{figure*}

% The example case - solving a maze using the left-hand rule
A toy example of a scenario-based model appears in
Fig.~\ref{fig:sbm-maze-example}. This model is designed to control a
\emph{Robotis Turtlebot 3} platform (\emph{Turtlebot}, for
short)~\cite{NaShVa21,AmSl19}. The
robot's goal is to perform mapless navigation towards a predefined
target, using information from lidar sensors and information about the
current angle and distance from the target. The scenarios are described as transition systems, where nodes represent synchronization points. The 
\j{MoveForward} scenario waits for the \j{InputEvent} event, which includes a 
payload vector, $v_t$, that contains sensor readings. If $v_t$ indicates that 
the area directly in front of the robot is clear, the scenario requests the 
event \j{Forward}. Clearly, in many cases moving forward is insufficient for 
solving a maze, and so we introduce a second scenario, \j{TurnLeft}. This 
scenario waits for an \j{InputEvent} event with a payload vector $v_t$ 
indicating that the area to the left of the robot is clear. It then requests 
the \j{Left} event. Further, the \j{TurnLeft} scenario blocks the \j{Forward}
event, to make the robot prefer a left turn to a move forward
(inspired by the \emph{left-hand rule}~\cite{wallFollower}). Finally,
The \j{MoveForward} scenario waits for the event \j{Left}, to return
to its initial state even if the \j{Forward} event was not triggered.

% Review of SBM
The SBM paradigm is well established, and has been studied thoroughly
in the past years. It has been implemented on top of Java~\cite{HaMaWe10}, JavaScript~\cite{BaWeRe18},
ScenarioTools, C++~\cite{HaKa14}, and Python~\cite{BPPy}; and has been used to model various
complex systems, such as cache coherence protocols, robotic
controllers, games, and more~\cite{HaKaMaMa16,AsMaWeWi15,HaKaMaMa18}. 
A key advantage of SBM is that its models can be checked and formally
verified~\cite{HaKaMaWe15}, and that automatic tools can be applied to
repair and launch SBM in distributed
environments~\cite{StGrGrHaKaMa18,HaKaMaWe14,HaKaLaMaWe15}.

In formalizing SBM, we follow the definitions of Katz~\cite{Ka13}. A
scenario object $O$ over a given event set $E$ is abstractly defined
as a tuple $O = \langle Q, q_0, \delta, R, B \ra$, where:

\begin{itemize}
	\item $Q$ is a set of states, each representing one of the
	predetermined synchronization points.
	\item $q_0\in Q$ is the initial state.
	\item $R:Q\to 2^E$ and $B:Q\to 2^E$ map states to the sets of
          events requested or blocked at these states (respectively).
	\item $\delta: Q \times E \to Q $ is a deterministic transition function, indicating how the scenario reacts when an event is triggered.
\end{itemize}

Let $M = \{O^1, ..., O^n\}$ be a be a behavioral model, where $n \in N$ and each $O^i = \langle Q^i, q_0^i, \delta^i, R^i, B^i\ra$ is a distinct scenario. In order to define the semantics of $M$, we construct a deterministic \emph{labeled transition system} $LTS(M) = \langle Q, q_0, \delta \ra$, where:
\begin{itemize}
	\item $Q := Q^1 \times ... \times Q^n$ is the set of states.
	\item $q_0 := \langle q_0^1, ..., q_0^n \ra \in Q $ is the
          initial state.
	\item $\delta: Q \times E \to Q$ is a deterministic transition function, defined for all $q = \langle q^1, ..., q^n \ra \in Q$ and $e \in E$, by:
\end{itemize}
\begin{equation*}
	\delta(q,e) := 
	\langle \delta^1(q^1,e), ...,\delta^n(q^n,e)\ra
\end{equation*}

An execution of $M$ is an execution of the induced $\lts(M)$. The
execution starts at the initial state $q_0$. In each
state $q=\langle q^1,\ldots,q^n\rangle \in Q$, the \emph{event selection mechanism (ESM)} inspects the set of
\emph{enabled events} $E(q)$ defined by:
\begin{equation*}
	\label{eq:e_q}
	E(q):=\bigcup\limits_{i=1}^{n} R^i(q^i) \setminus \bigcup\limits_{i=1}^{n} B^i(q^i)
\end{equation*}

If $E(q) \neq \emptyset$, the mechanism selects an event $e\in E(q)$
(which is requested and not blocked). Event $e$ is then triggered, and
the system moves to the next state, $q'=\delta(q,e)$, where the
execution continues. An execution can be formally recorded as a
sequence of triggered events, called a \emph{run}. The set of all
\emph{complete} runs is denoted by
$\mathcal{L}(M) \triangleq \mathcal{L}(\lts(M))$. It contains both
infinite runs, and finite runs that end in \emph{terminal states},
i.e.~states in which there are no enabled events.

\subsection{Modeling Override Scenarios using SBM}
\label{sec:modeling-override-scenarios}

\begin{figure}[ht]
  \centering
  \includegraphics[scale=0.4]{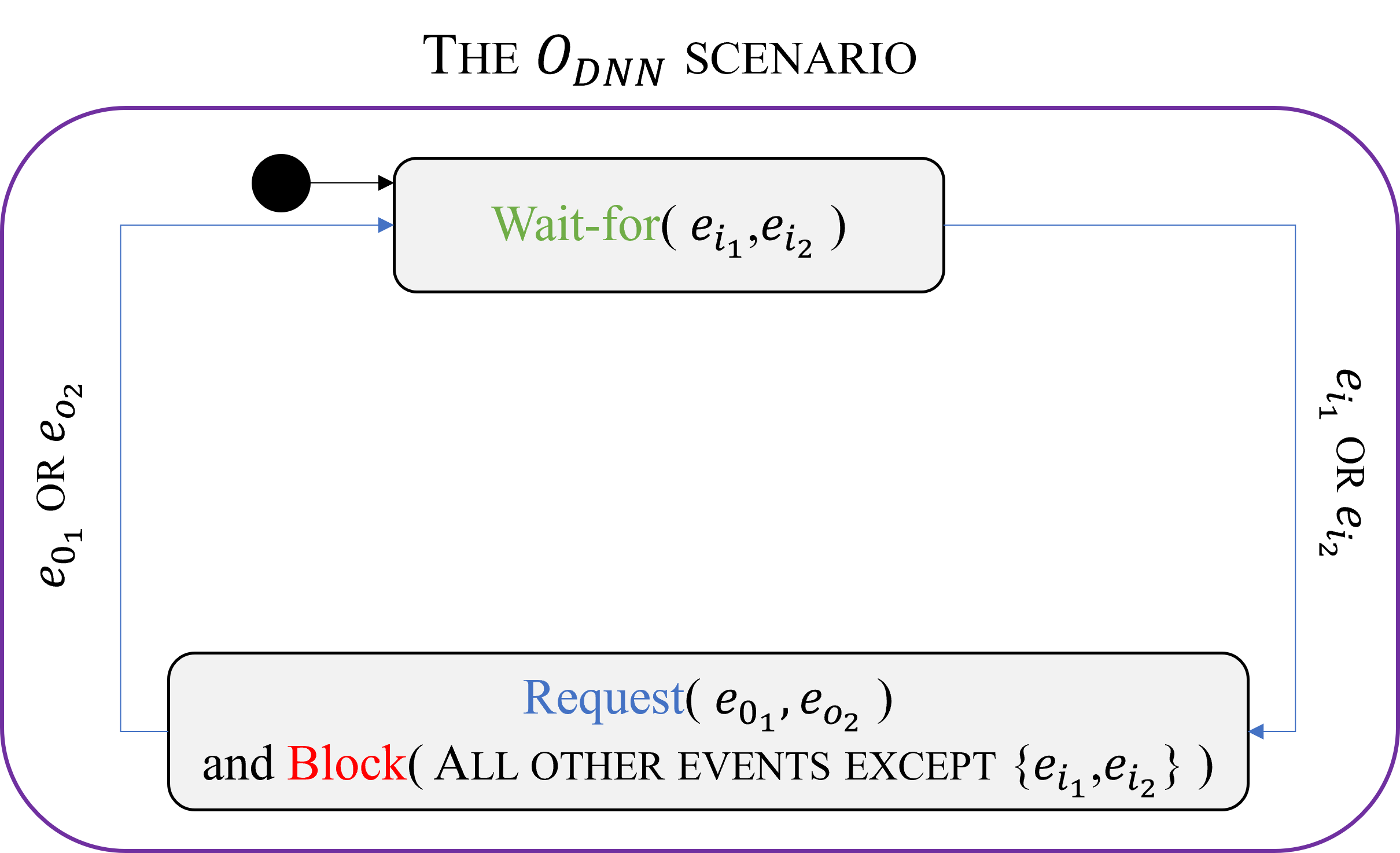}  
  \caption{A figure of the $O_{DNN}$ scenario object corresponding to the
  	neural network in Fig.~\ref{fig:dnn_running_example} described in  
  	statecharts. The black circle indicates the initial state. The scenario 
  	waits for the events $e_{i_1}$ and $e_{i_2}$ that represent the inputs to the neural network. These events contain a payload with the actual values 
  	assigned. The scenario then proceeds to request the events $e_{o_1}$ and
    $e_{o_2}$, which represent the possible output labels $y_1, y_2$ 
    respectively (inspired by~\cite{Ka20b}).}
  \label{fig:scenarioObject}
\end{figure}

% The ODNN design pattern
We follow a recently proposed
method~\cite{Ka20a,Ka20b} for designing SBM
models that integrate scenario objects and a DNN controller. The main
concept is to represent the DNN as a scenario object, $O_{DNN}$, that
operates as part of the scenario-based model, enabling the different
scenarios to interact with the DNN. As a first step, we assume that
there is a finite set of possible inputs to the DNN, denoted
$\mathbb{I}$; and let $\mathbb{O}$ mark the set of possible actions
the DNN can select from (we relax the limitation of finite event sets
later on). We add new events to the event set $E$: an event $e_i$ that contains 
a payload of the input values for every $i\in \mathbb{I}$, and an event $e_o$ 
for every $o\in \mathbb{O}$. The scenario object $O_{DNN}$ continually waits for
all events $e_i$, and then requests all output events $e_o$. This modeled
behavior captures the black-box nature of the DNN: after an input
arrives, one of the possible outputs is chosen, but we do not know
which. However, when the model is deployed, the execution
infrastructure evaluates the actual DNN, and triggers the event that
it selects. For instance, assuming that there are only two possible
inputs: $i_1 = \langle 1,0\rangle$ and $i_2 = \langle 0,1\rangle$, the
network portrayed in Fig.~\ref{fig:dnn_running_example} would be
represented by the scenario object depicted in
Fig.~\ref{fig:scenarioObject}.

% ODNN high level infrastructure 
By convention, we stipulate that scenario objects in the system may
wait-for the input events $e_i$, but may not block them. A
dedicated scenario object, the \emph{sensor}, is in charge of
requesting an input event when the DNN needs to be evaluated. Another
convention is that only the $O_{DNN}$ may request the output events, $e_o$;
although other scenarios may wait-for or block these events. At
run time, if the DNN's classification result is an event which is
currently blocked, the event selection mechanism resolves this by
selecting a different output event which is not blocked. If there are
no unblocked events, the system is considered deadlocked, and the
SBM program terminates. The motivation for these conventions is to allow
scenario objects to monitor the DNN's inputs and outputs. The
scenarios can then intervene, and override the DNN's output --- by
blocking specific output events. An override scenario can coerce the
DNN to select a specific output, by blocking all other output events;
or it can interfere in a more subtle manner, by blocking some output
events, while allowing $O_{DNN}$ to select from the remaining ones.
One strategy for selecting an alternate output event in a classification problem will be to select the event with the next-to-highest score.

In practice, the requirement that the event sets $\mathbb{I}$ and
$\mathbb{O}$ be finite is restrictive, as DNNs typically have a very
large (effectively infinite) number of possible inputs.  To overcome
this restriction, we follow the extension proposed
in~\cite{KaMaSaWe19}, which enables us to treat events as typed
variables, or sets thereof. Using this extension, the various scenarios
can affect, through requesting and blocking, the possible values of
these variables; and a scenario object's transitions may be
conditioned upon the values of these variables. In particular, these
variables can be used to express an infinite  number of possible inputs and
outputs of a DNN.

\begin{figure}[ht]
	\centering
	\includegraphics[scale=0.33]{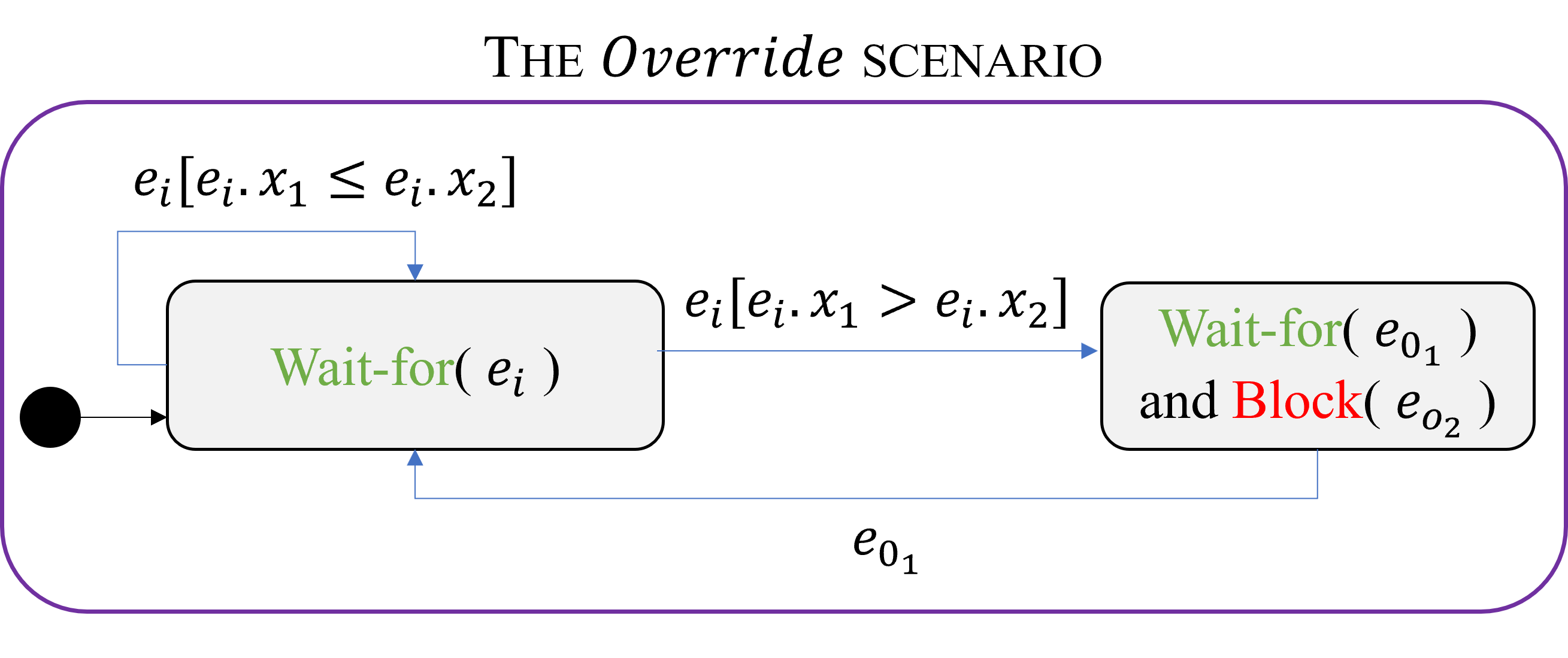}  
	\caption{A scenario object for enforcing the override rule defined in
  	Sec.~\ref{sec:background-overrideRules}. The scenario waits for the input 
  	event $e_i$ and inspects the payload to see if the predicate $P$ holds for 
  	the given input. It then continues to wait for the output event $e_{o_1}$, 
  	while blocking the unwanted event $e_{o_2}$. This blocking
        forces the triggering of output event $e_{o_1}$. Once this happens, the 
  	scenario returns to its initial state.}
  \label{fig:overrideRuleAsThread}
\end{figure}

Using the aforementioned extension, the override rule from
Sec.~\ref{sec:background:dnns} is depicted in
Fig.~\ref{fig:overrideRuleAsThread}. The
scenario waits for the input event $e_i$, which now contains as a
payload two real-valued variables, $x_1$ and $x_2$, that represent
the actual assignment to the DNN's inputs. The transitions of the
scenario object are then conditioned upon the values of these
variables: if the predicate $P$ holds for this input, the scenario
transitions to its second state, where it overrides the DNN's output
by blocking the output event $e_{y_{2}}$, which necessarily causes the
triggering of $e_{y_{1}}$.

\section{\uppercase{Case Study: The Aurora Congestion Controller}}
\label{sec:case_aurora}

% High level description
For our first case study, we focus on \emph{Aurora}~\cite{JaRoGoScTa19}
--- a recently proposed \emph{performance-oriented congestion control
  (PCC)} protocol, whose purpose is to manage a computer network
(e.g., the Internet). Aurora's goal is to maximize the network's
throughput, and to prevent ``congestive collapses'', i.e., situations
where the incoming traffic rate exceeds the outgoing bandwidth and
packets are lost. Aurora is powered by a DRL-trained DNN agent that
attempts to learn an optimal policy with respect to the environment's
\emph{state} and \emph{reward}, which reflect the agent's performance
in previous batches of sent packets. The \emph{action} selected by
the agent is the sending rate that is used for the next batch of
packets. It has been shown that Aurora can obtain impressive results,
competitive with modern, hand-crafted algorithms for similar
purposes~\cite{JaRoGoScTa19}.

% Specific description of Aurora's input/output
Aurora employs the concept of \emph{monitor
  intervals (MIs)}~\cite{DoMeZaArGiGoSc18}, in which time is split into
consecutive intervals. At the start of each MI, the agent's chosen
action $a_t$ (a real value) is selected as the sending rate for the
current MI, and it remains fixed throughout the interval. This rate
affects the pace, and eventually the throughput, of the
protocol. After the MI has finished, a vector $v_t$ containing
real-valued performance statistics is computed from data
collected during the interval. Subsequently, $v_t$ is provided as the
environment state to the agent, which then proceeds to select a new sending rate
$a_{t+1}$ for the next MI, and so on. For a more extensive background
on performance-oriented congestion control, see~\cite{DoLiZaGoSc15}.

% Simulator and starting point
As a supporting tool, Aurora is distributed with the \emph{PCC-DL} simulator~\cite{PCCUspaceDL} that
enables the user to test Aurora's performance. The simulator has two
built-in congestion control protocols:
\begin{itemize} 
	\item The \emph{PCC-IXP} protocol: a simple protocol that adjusts the
	sending rate using a hard-coded function.
	\item The \emph{PCC-Python} protocol: a protocol that utilizes a trained	Aurora agent to adjust the sending rate.
\end{itemize}
Both of these protocols are classified as normal \emph{(primary)}
protocols that aspire to maximize their throughput~\cite{MeScGoSc20}.

% The reason we selected Aurora as a case study.
We chose Aurora as our first case study because of its reactive nature: it 
receives external input from the environment, processes this information using 
the trained DRL agent, and acts on it with the next sending rate. SBM is well 
suited for reactive systems~\cite{HaMaWe12}, and Aurora matched our 
requirements to enhance a reactive DL system. The goals we set out to achieve 
in this case study are detailed in the following section.

\subsection{Integrating Aurora and SBM}
\label{sec:case_aurora_integration_sbm}

% Aurora extension - first goal
Our first goal was to instrument the Aurora DNN agent with the
$O_{DNN}$ infrastructure, and integrate it with an SB model. This was
achieved through the inclusion of the C++ SBM package~\cite{HaKa14,BPC} in
the simulator; and the introduction of a new protocol, \emph{PCC-SBM},
which extends the PCC-Python protocol and launches an SB model that
includes the $O_{DNN}$ scenario. This process, on which we elaborate
next, required significant technical work --- and successfully
produced an integrated SBM/DNN model that performed on par with the
original, DNN-based model.

\begin{figure}[htp]
	\centering
	\includegraphics[scale=0.5]{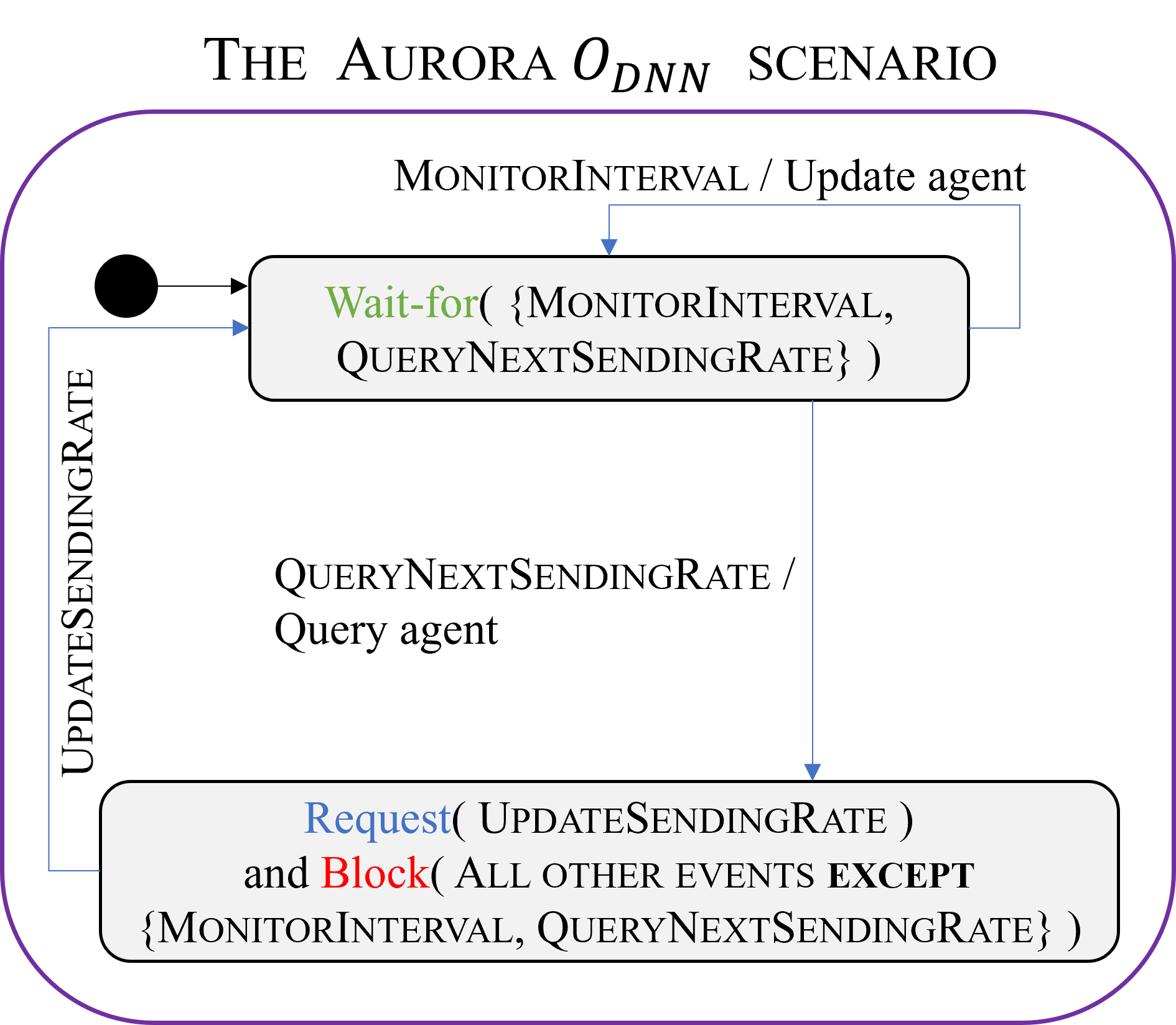}	
	\caption{The \j{Aurora} \j{$O_{DNN}$} scenario.}
	\label{fig:pcc-sbm-odnn}
\end{figure}

% Description - Sensors, Actuator, O_DNN scenario - input/output events
The simulator interacts with the PCC-SBM protocol in two ways:
\begin{inparaenum}[(i)]
\item it provides the statistics of the current MI; and
\item it requests the next sending rate.
\end{inparaenum}
Thus, we began the SBM/DNN integration by introducing a \j{Sensor}
scenario, whose purpose is to inject \j{MonitorInterval} and
\j{QueryNextSendingRate} events into the SB model, to allow
it to communicate with the simulator. Fig.~\ref{fig:pcc-sbm-odnn}
depicts the \j{Aurora $O_{DNN}$} scenario (using a combination of
Statecharts and SBM visual
languages~\cite{Ha86,MaHaHaMlTe18}), which waits for
these events in its initial state. The event \j{MonitorInterval}
carries, as a payload, the MI statistics vector, $v_t$, whose entries
are real values.

When \j{MonitorInterval} is triggered, the statistics vector is
provided as the input to the underlying DNN, and when the event
\j{QueryNextSendingRate} is triggered, the scenario extracts the DNN's
output, and then uses it as a payload for an \j{UpdateSendingRate}
that it requests --- while blocking all other, non-input events.
Finally, we introduce an \j{Actuator} scenario, which waits for the
event \j{UpdateSendingRate} and updates the simulator on the selected
sending rate for the next batch of packets. 

\subsection{Supporting Scavenger Mode}
\label{sec:scavengerMode}

% Aurora extension - secondary goal
For our second, more ambitious goal, we set out to extend the Aurora
system with a new behavior, without altering its underlying DNN: specifically,
with the ability to support \emph{scavenger mode}~\cite{MeScGoSc20}. The 
scavenger protocol is a ``polite'' protocol, meaning it can yield its 
throughput if there is competition in the same physical network. Of course, 
such behavior needs to be temporary, and when the other traffic on the physical 
network subsides, the scavenger protocol should again increase its throughput, 
utilizing as much of the available bandwidth as possible.

In order to add scavenger mode support, we added the following scenarios:
\begin{itemize}
\item The \j{MonitorNetworkState} scenario object, which inspects the
  state of the physical network and requests a specific event: the
  \j{EnterYield} event that marks that the conditions for
  entering \emph{yield mode}, in which sending rates should be
  reduced, are met.
\item The \j{ReduceThroughput} scenario object, which is an override
  scenario. This scenario first waits-for a notification that the
  protocol should enter yield mode, and then proceeds to override the
  DNN's calculated sending rate with a lower sending rate.
\end{itemize}
Our plan was for the \j{ReduceThroughput} scenario  to support three override policies:
\begin{inparaenum}[(i)]
\item an immediate decline to a fixed, low sending rate; 
\item a gradual decline, using a step function; and
\item a gradual decline, using exponential decay.
\end{inparaenum}
However, we quickly observed that the existing override scenario
formulation (as presented in
Sec.~\ref{sec:modeling-override-scenarios}) was not suitable for this
task.

Recall that an override scenario overrides $O_{DNN}$'s output by
blocking any unwanted output events, and coercing the event selection
mechanism to select a different output event that is not blocked.
In our case, however, we needed \j{ReduceThroughput} to act as an
override scenario that blocks some output events based on the output selected
by $O_{DNN}$, in the \emph{current time step}. For example, in the
case of a gradual decline in the sending rate, if $O_{DNN}$ would
normally select sending rate $x$, we might want to force the selection
of rate $\frac{x}{2}$, instead; but this requires knowing the value of
$x$, in advance, which is simply not possible using the current
formulation~\cite{Ka20a}.

\begin{figure*}[htp]
	\centering
	\includegraphics[scale=0.4]{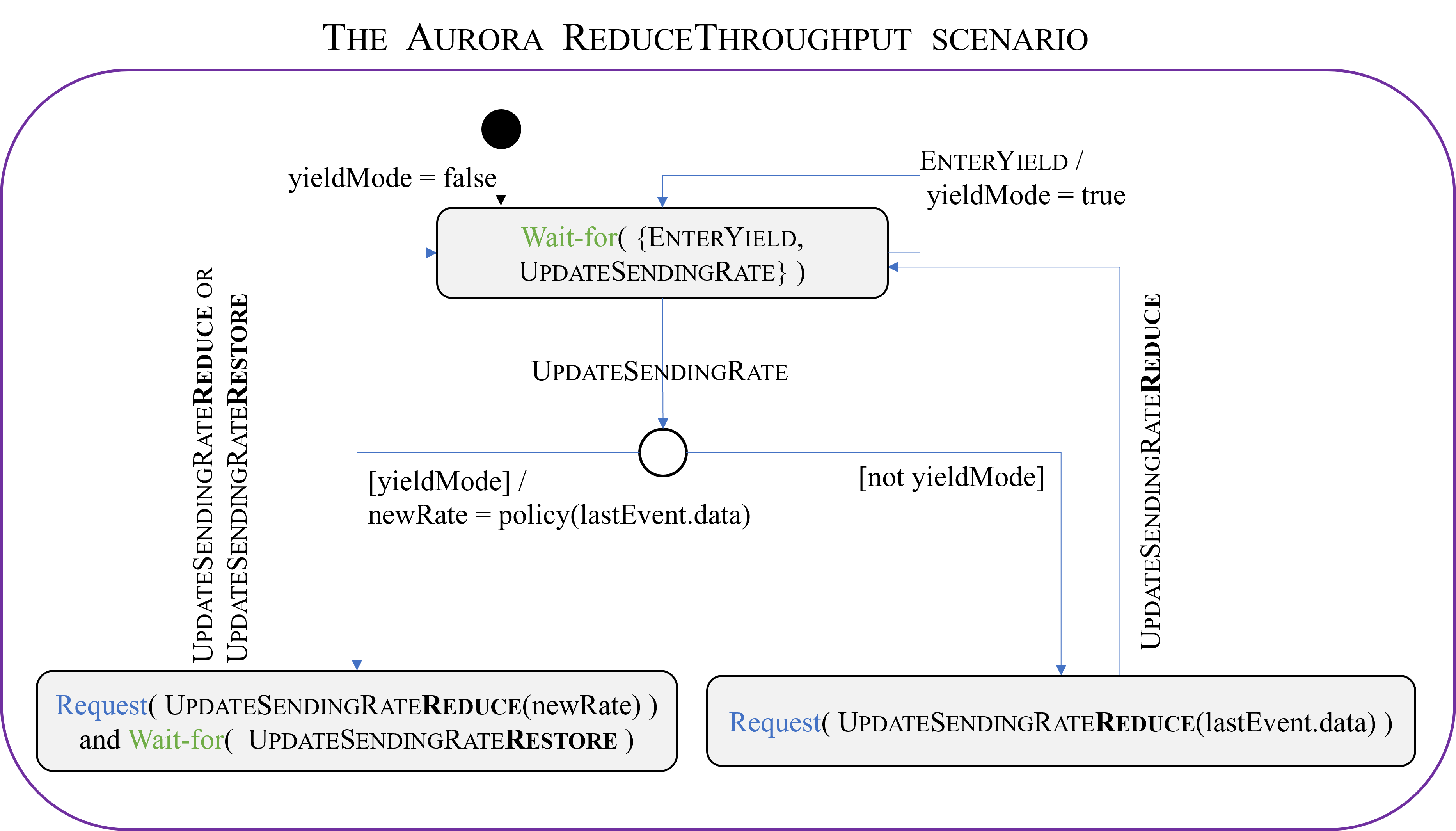}	
	\caption{The Aurora \j{ReduceThroughput} scenario, described using Statecharts enhanced with SBM. The black circle specifies the initial state. The scenario waits-for \j{EnterYield} event to enter yield mode. If yield mode is enabled, the \j{UpdateSendingRate} event payload will be modified, and the \j{UpdateSendingRateReduce} will be requested. Otherwise, the payload is propagated as-is in the ``proxy'' \j{UpdateSendingRateReduce}  event. The scenario waits for \j{UpdateSendingRateRestore} to return to its initial state, in case \j{UpdateSendingRateReduce} is blocked.}
	\label{fig:pcc-sbc-reduce}
\end{figure*}

To circumvent this issue within the existing modeling framework, we
introduce a new ``proxy event'', \j{UpdateSendingRateReduce}, intended
to serve as a middleman between the \j{Aurora $O_{DNN}$} scenario and
its consumers.  Our override scenario, \j{ReduceThroughput}, no longer
directly blocks certain values that the DNN might produce. Instead, it
waits-for the \j{UpdateSendingRate} event produced by \j{Aurora
  $O_{DNN}$}, manipulates its real-valued payload as needed, and then
requests the proxy event \j{UpdateSendingRateReduce} with the
(possibly) modified value. Then, in every scenario that originally
waited-for the \j{UpdateSendingRate} event, we rename the event to
\j{UpdateSendingRateReduce}, so that the scenario now waits for the
proxy event, instead. Fig.~\ref{fig:pcc-sbc-reduce} visually illustrates the
final version of the \j{ReduceThroughput} scenario.

% Restore throughput scenario
After entering scavenger mode and lowering the sending rate, a natural
requirement is that the system eventually reverts to a higher sending
rate, when scavenger mode is no longer required. To achieve this, we
adjust the \j{MonitorNetworkState} scenario to dynamically identify
this situation, and signal to the other scenarios that the system has
entered \emph{restore} mode, by requesting the event \j{EnterRestore}. We 
then introduce a second override scenario, \j{RestoreThroughput}, that can 
increase the protocol's throughput according to one of two predefined policies:
(i) an immediate return to the model's original output; 
or (ii) a \emph{slow start} policy~\cite{SlowStart}.

The \j{RestoreThroughput} scenario waits-for the events
\j{EnterRestore}, \j{EnterYield} and \j{UpdateSendingRate}.
The first two events signal the scenario to enter/exit restore
mode. When \j{UpdateSendingRate} is triggered and the scenario is in
restore mode, it overrides the value according to the policy in use,
and requests an output event with a modified value. Utilizing the
\j{UpdateSendingRateReduce} event for this purpose would result in
two, likely contradictory output events being requested at a single
synchronization point. To avoid this, we introduce a new event, \j{UpdateSendingRateRestore}, to be requested by the
\j{RestoreThroughput} scenario, while blocking the possible
\j{UpdateSendingRateReduce} event at the synchronization point. This
decision prioritizes ratio restoration over yielding (although any
other prioritization rule could be used). Finally, in every scenario
that requests/waits-for the \j{UpdateSendingRateReduce} event, we add
a wait-for the \j{UpdateSendingRateRestore} event. In this manner,
these scenarios can proceed with their execution despite being
blocked.

\subsection{Evaluation}

% Overview of accomplishments
For evaluation purposes, we implemented the scenario objects described
in Sec.~\ref{sec:scavengerMode}, and then used Aurora's simulator to
evaluate the enhanced model's performance, compared to that of the
original~\cite{SBMEnhanceDL}. Our results, described
below, indicate that the modified system successfully supports
scavenger mode, although its internal DNN remained unchanged.

\begin{figure}
	\centering
	\includegraphics[width=1.0\linewidth]{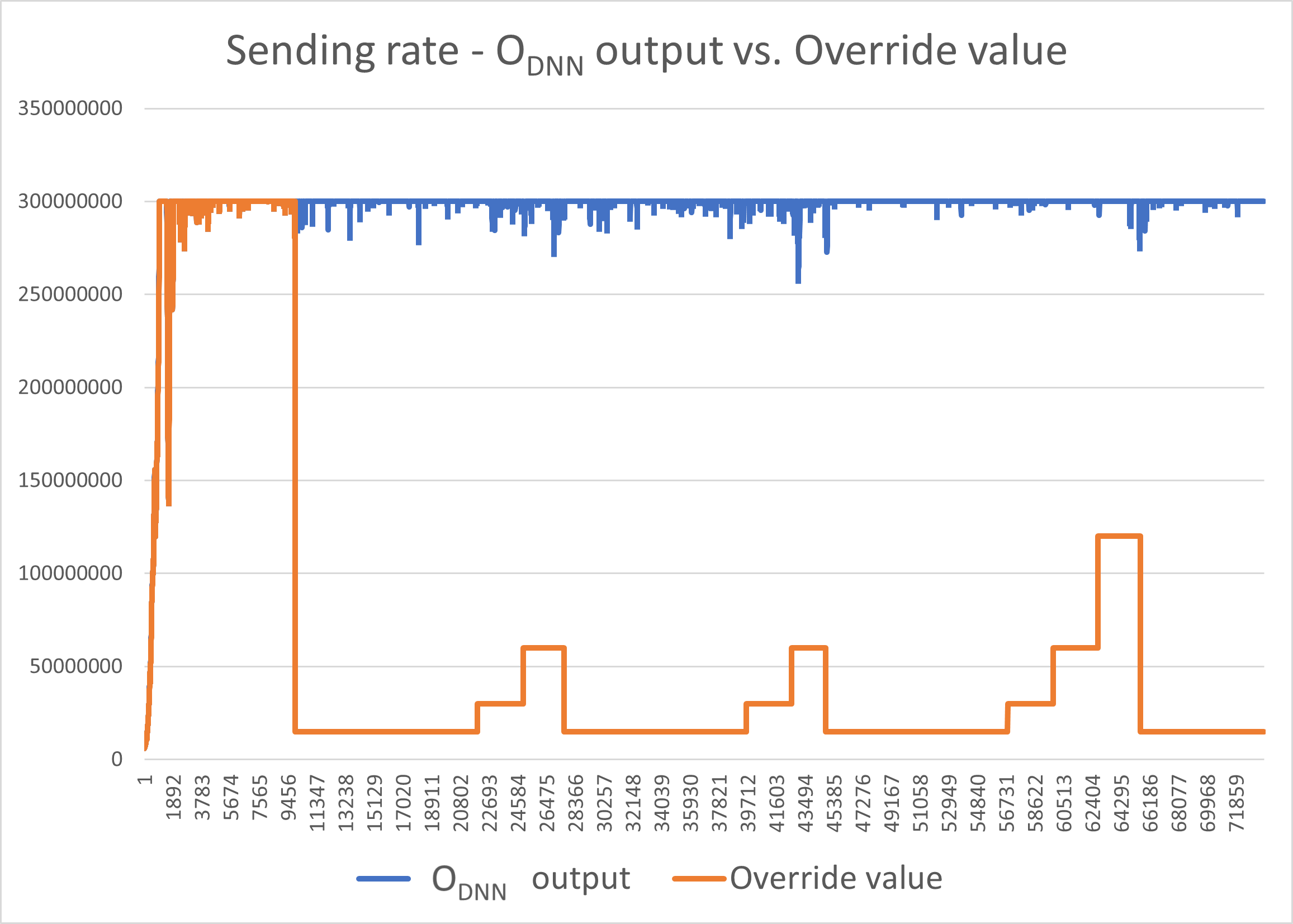}
	\caption{The \j{Aurora $O_{DNN}$} original model output,
          vs.~the values produced by the override scenarios. The policy used for
          reduction is an immediate decline to a fixed rate. The
          policy used for restoration is ``slow start''.}
	\label{fig:send-rate-yield-and-restore}
\end{figure}

% Describe model output vs overriden value
Fig.~\ref{fig:send-rate-yield-and-restore} depicts the sending rate
requested by the \j{Aurora $O_{DNN}$} scenario, following an input
event \j{QueryNextSendingRate}, and the actual sending rate that was
eventually returned to the simulator by the PCC-SBM protocol. We
notice that initially, the two values coincide, indicating that no
overriding is triggered --- because the \j{MonitorNetworkState}
scenario did not yet signal that the system should enter yield
mode. However, once this signal occurs, the \j{ReduceThroughput}
scenario overrides the sending rate, according to the fixed rate
policy. After a while, the \j{MonitorNetworkState} detects that it is
time to once again increase the sending rate, and signals that the
system should enter restore mode. As a result, we see an increase, per
the ``slow start'' restoration policy of \j{RestoreThroughput}. The
ensuing back-and-forth switching between yield and restore modes
demonstrates that the \j{MonitorNetworkState} scenario dynamically
responds to changes in environmental conditions.

\begin{figure}
	\centering
	\includegraphics[width=1.0\linewidth]{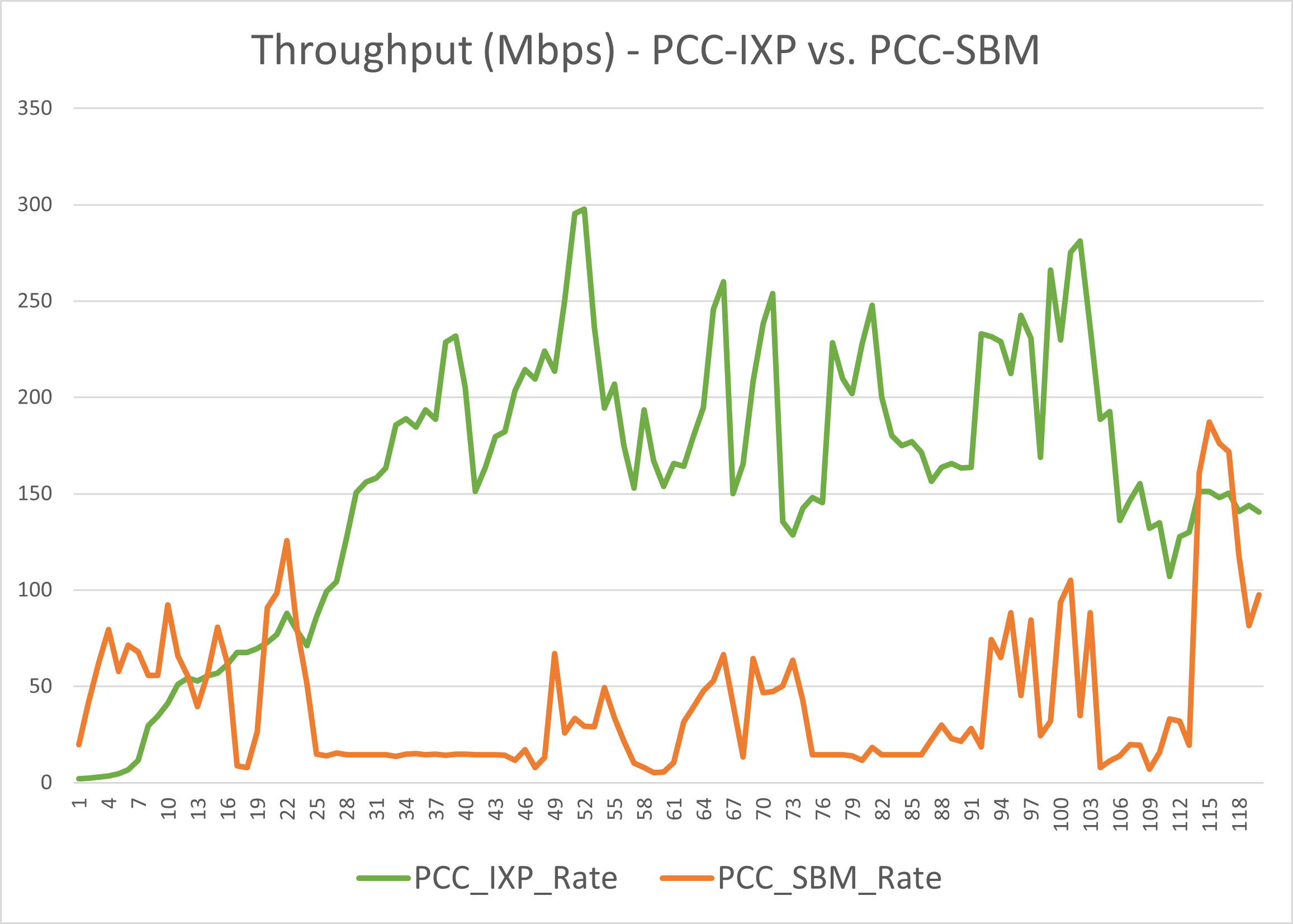}
	\caption{The recorded throughput (MB/s) of the PCC-IXP and
          PCC-SBM protocols, when executed in parallel using the PCC
          simulator.}
	\label{fig:throughput_pcc-ixp_vs_pcc-sbm}
\end{figure}

% Experiment - PCC-IXP vs. PCC-SBM
In another experiment, we compared the throughput (MB/s) of the
primary PCC-IXP protocol with that of the PCC-SBM protocol, when the
two are executed in parallel. The results appear in
Fig.~\ref{fig:throughput_pcc-ixp_vs_pcc-sbm}.  We observe that there
is a resemblance between the overridden sending rate value seen in
Fig.~\ref{fig:send-rate-yield-and-restore} and the actual
throughput. When the \j{MonitorNetworkState} scenario signals to
yield, the sending rate declines to a fixed value, which in turn leads
to a fixed throughput. Additionally, when the sending rate increases
after a signal to restore, the throughput of the protocol increases as
well. Another interesting phenomenon is that when the PCC-SBM
relinquishes bandwidth, the PCC-IXP increases its own throughput,
which is the behavior we expect to see. We speculate that the yield
of the PCC-SBM enabled this increase.

\section{\uppercase{Case Study: The Robotis Turtlebot3 platform}}
\label{sec:case_robotis}

% Introduction to TurtleBot and mapless maze naviagation
For our second case study, we chose to enhance a DL system trained by Corsi et 
al.~\cite{CoYeAmFaHaKa22,ScenarioBasedRL}, which aims to solve a setup of
the \emph{mapless navigation} problem. The system contains a DNN
agent whose goal is to navigate a \emph{Turtlebot 3 (Turtlebot)} 
robot~\cite{TurtleBot3,AmSl19} to a target destination, without 
colliding with obstacles. Contrary to classical planning, the robot
does not hold a global map, but instead attempts to navigate using
readings from its environment. A successful navigation policy must thus be 
dynamic, adapting to changes in local observations as the robot moves closer to 
its destination. DRL algorithms have proven successful in learning such a 
policy~\cite{MaFa20}.

% Specific description of Turtlebot  - Input and Output events
We refer to the DNN agent that controls the Turtlebot as 
TRL (for Turtlebot using RL). The agent learns a navigation policy iteratively: in each iteration, it is provided with a vector $v_t$ that comprises
\begin{inparaenum}[(i)]
\item normalized lidar scans of the robot's distance from  any
  nearby obstacles; and
\item the angle and the distance of the robot from the target.
\end{inparaenum}
The agent then evaluates its internal DNN on $v_t$,
obtaining a vector $v_{a_t}$ that contains a probability distribution
over the set of possible actions the Turtlebot can perform: moving
forward, or turning left or right. For example, one possible vector is
$ v_{a_t} = [(\text{Forward}\xspace{}, 0.2), (\text{Left}\xspace{},0.5), (\text{Right}\xspace{},0.3)]$. Using this
vector, the agent then randomly selects an action according to the
distribution, navigates the Turtlebot, and receives a reward.

% Simulation of the TRL in unity.
The DNN at the core of the TRL controller is trained and tested in a simulation
environment that contains a sim-robot Turtlebot 3 burger~\cite{TurtleBot3}, and a single, 
fixed maze, created using the \emph{ROS2} framework~\cite{ROS2} and the Unity3D 
engine~\cite{UnityAndRos}. In each session, the robot's starting location is
drawn randomly, which enables a diverse scan of the input space. The navigation
session has four possible 
outcomes:
\begin{inparaenum}[(i)]
\item success;
\item collision;
\item timeout; or
\item an unknown failure.
\end{inparaenum}

% The reason we selected Turtlebot as a case study.
We selected the Turtlebot project as our second case study due to its reactive 
characteristics: it reads external information using its sensors, applies an 
internal logic to select the next action, and then acts 
by moving towards the target. SBM has previously been applied to model robot 
navigation and maze solving~\cite{El21,AsGoMaStWe17}, which 
strengthened our intuition that an enhancement of the Turtlebot with SBM is 
feasible. Next, we outline the objectives we aimed to achieve in this case study.

\subsection{Integrating Turtlebot and SBM}

% Application of the ODNN pattern in the TRL case
Similarly to the Aurora case, our first goal was to instrument the
Turtlebot DNN with the $O_{DNN}$ infrastructure, so that it could be
composed with an SB model. This was achieved by using the Python
implementation of SBM~\cite{BPPy}, and integrating it with the TRL
code. We created a \j{Sensor} scenario that waits for the current
state vector $v_t$, and injects an \j{InputEvent} containing $v_t$
into the SB model; and also an \j{Actuator} scenario that waits for
an internal \j{OutputEvent}, and transmits its action as the one to be
carried out by the Turtlebot.

\begin{figure}[ht]
	\centering
	\includegraphics[width=0.8\linewidth]{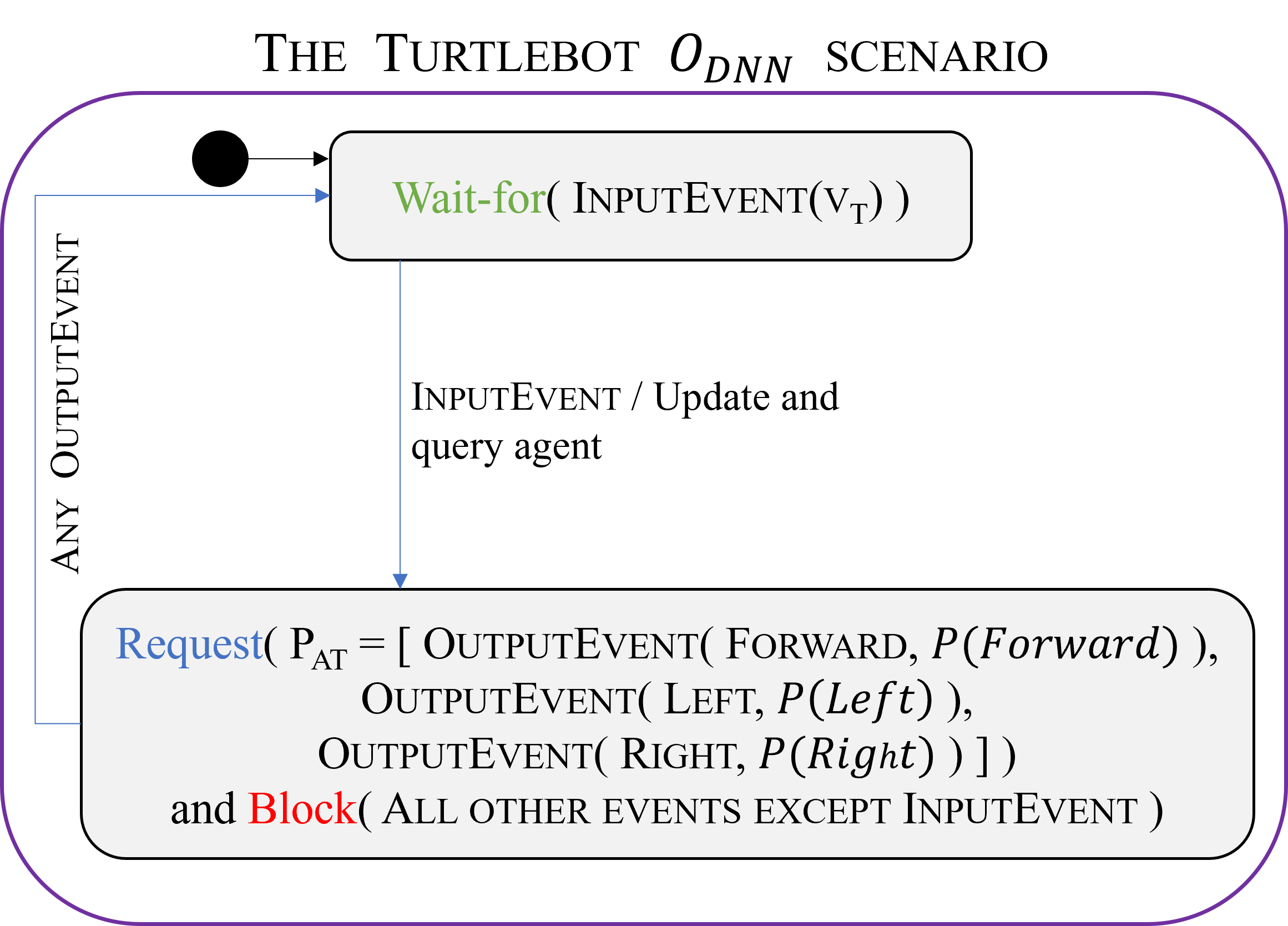}
	\caption{The \j{Turtlebot $O_{DNN}$} scenario. The scenario
          waits for an \j{InputEvent} containing $v_t$, provides it to
          the agent, and receives a vector $P_{at}$ of possible
          actions and probabilities. It then proceeds to request all
          the output events from the ESM using $P_{at}$. At the synchronization point, the ESM executes the \j{Turtlebot $O_{DNN}$} event selection strategy, and one possible \j{OutputEvent} is triggered.}
		\label{fig:turtlebot-odnn}
\end{figure}

% The Robotis ODNN
Next, we proceeded to create the \j{Turtlebot $O_{DNN}$} scenario
for TRL.  Unlike in the Aurora case, where the DNN would output a
single chosen event, here the DNN outputs a probability distribution
over the possible actions (a common theme in
DRL-based systems~\cite{SuBa18}). To
accommodate this, we adjusted our $O_{DNN}$ scenario to request
all the possible output events in the form of a vector $P_{at}$, which 
contains each possible action and its probability. We then 
modified SBM's event selection mechanism to randomly select a requested output 
event from $P_{at}$, with respect to the induced probability distribution. The 
mechanism then triggers the \j{OutputEvent}, which contains in its payload the 
selected action and its probability. During modeling and 
experimentation, the scenario can assign any probability distribution (e.g., 
uniform) to the DNN's output events; and during 
deployment, these values are computed using the actual DNN (see 
Fig.~\ref{fig:turtlebot-odnn}). If the event
selection mechanism selects an event that is blocked, the selection is
repeated, until a non-blocked event is selected.  If there are no
enabled events, then the system is deadlocked and the program
ends. Using this approach, our Turtlebot controller could successfully
navigate in various mazes.

% ODNN figure

Once the $O_{DNN}$ infrastructure was in place, we verified that the
augmented program's performance was similar to that of the original
agent. This was achieved by comparing the models pre-trained by~\cite{CoYeAmFaHaKa22} to our SBM-enhanced version, and
checking that both agents obtained similar success rates on various
mazes.

\subsection{A Conservative Controller}
\label{sec:case_robotis_conservative}

\begin{figure}[ht]
	\centering
	\includegraphics[width=0.8\linewidth]{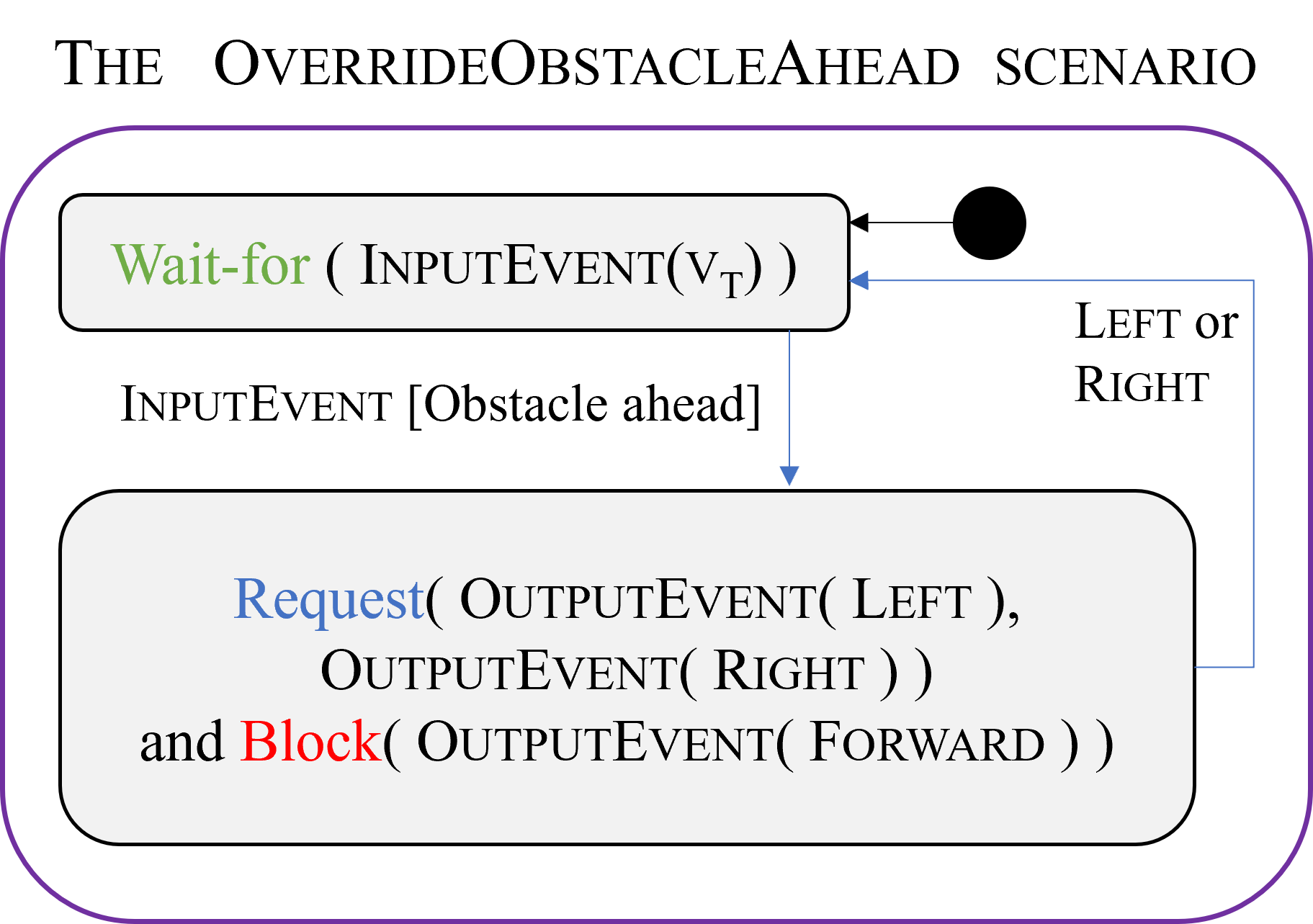}
	\caption{The \j{OverrideObstacleAhead} scenario waits for an \j{InputEvent} containing $v_t$, and inspects the lidar sensors facing forward readings. If $distance<0.22$, moving forward will cause a collision. Therefore the scenario blocks the \j{OutputEvent(Forward)} event.} 
	\label{fig:turtlebot-basic-guard}
\end{figure}

For our second goal, we sought to increase the model's safety, by
implementing a basic override rule, \j{OverrideObstacleAhead}, which would 
prevent the Turtlebot from colliding with an obstacle that is directly ahead. 
This can be achieved by analyzing the DNN's inputs, which include the lidar 
readings, and identifying cases where a move forward would guarantee a 
collision; and then blocking this move, forcing the system to select a different
action. An illustration of this simple override rule appears in
Fig.~\ref{fig:turtlebot-basic-guard}.

% The motivation for the conservative guard rule
As we were experimenting with the Turtlebot and various override
rules, we noticed the following, interesting pattern. Recall that a
Turtlebot agent learns a policy, which, for a given state $s_t$,
produces a probability distribution over the actions,
$a_t=[P(\text{Forward}\xspace{}), P(\text{Left}\xspace{}), P(\text{Right}\xspace{})]$. We can regard this vector as the agent's 
\emph{confidence} levels that each of the possible actions will bring the 
Turtlebot closer to its goal. The random selection that follows takes these 
confidence levels into account, and is more likely to select an action that the
agent is confident about; but this is not always the
case. Specifically, we observed that for ``weaker'' models,
e.g.~models with about a 50\% success rate, the agent would repeatedly
select actions with a \emph{low} confidence score, which would often
lead to a collision. This observation led us to design our next
override scenario, \j{ConservativeAction}, which is intended to force
the agent to select actions only when their confidence score meets a
certain threshold.

% The conservative rule
% The repeated pattern of the proxy event - and modifications
Ideally, we wish for \j{ConservativeAction} to implement the following behavior:
\begin{inparaenum}[(i)]
	\item wait-for the \j{OutputEvent} being selected
	\item examine whether the confidence score in its payload is below a certain threshold, and if so,
	\item apply blocking to ensure that a different \j{OutputEvent}, with a higher score, is selected for triggering.
\end{inparaenum}
This method again requires that the override scenario be able to inspect the 
content of the \j{OutputEvent} being triggered in the current time step. 

To overcome this issue, we add a new, proxy event called
\j{OutputEventProxy}, and adjust all existing scenarios that would
previously wait for \j{OutputEvent} to wait for this new event,
instead.  Then, we have the \j{ConservativeAction} scenario wait-for
the input event to $O_{DNN}$, and have it \emph{replay} that event to
initiate additional evaluations of the DNN, and the ensuing random
picking of the \j{OutputEvent}, until an acceptable output action is
selected. When this occurs, the \j{ConservativeAction} scenario
propagates the selected action as an \j{OutputEventProxy} event.

\subsection{Evaluation}

For evaluation purposes, we trained a collection of agents, $C_{agents}$, using the
technique of Corsi et al.~\cite{ScenarioBasedRL}. These agents had varying success rates,
ranging from $4\%$ to $96\%$. Next, we compared the performance of these
agents to their performance when enhanced by our SB model.

In the first experiment, we disabled our override rule, and had 
every agent in $C_{agents}$ solve a maze from 100 different random
starting points. The statistics we measured were:
\begin{itemize}
\item num\_of\_solved: the number of times the agent reached the target.
\item num\_of\_collision: the number of times the agent collided with an obstacle.
\item avg\_num\_of\_steps: the average number of steps the agent
  performed in a successful navigation.
\end{itemize}
We then repeated this setting with the \j{ConservativeAction}
scenario enabled.

\begin{figure}[ht]
	\centering
	\includegraphics[width=1.0\linewidth]{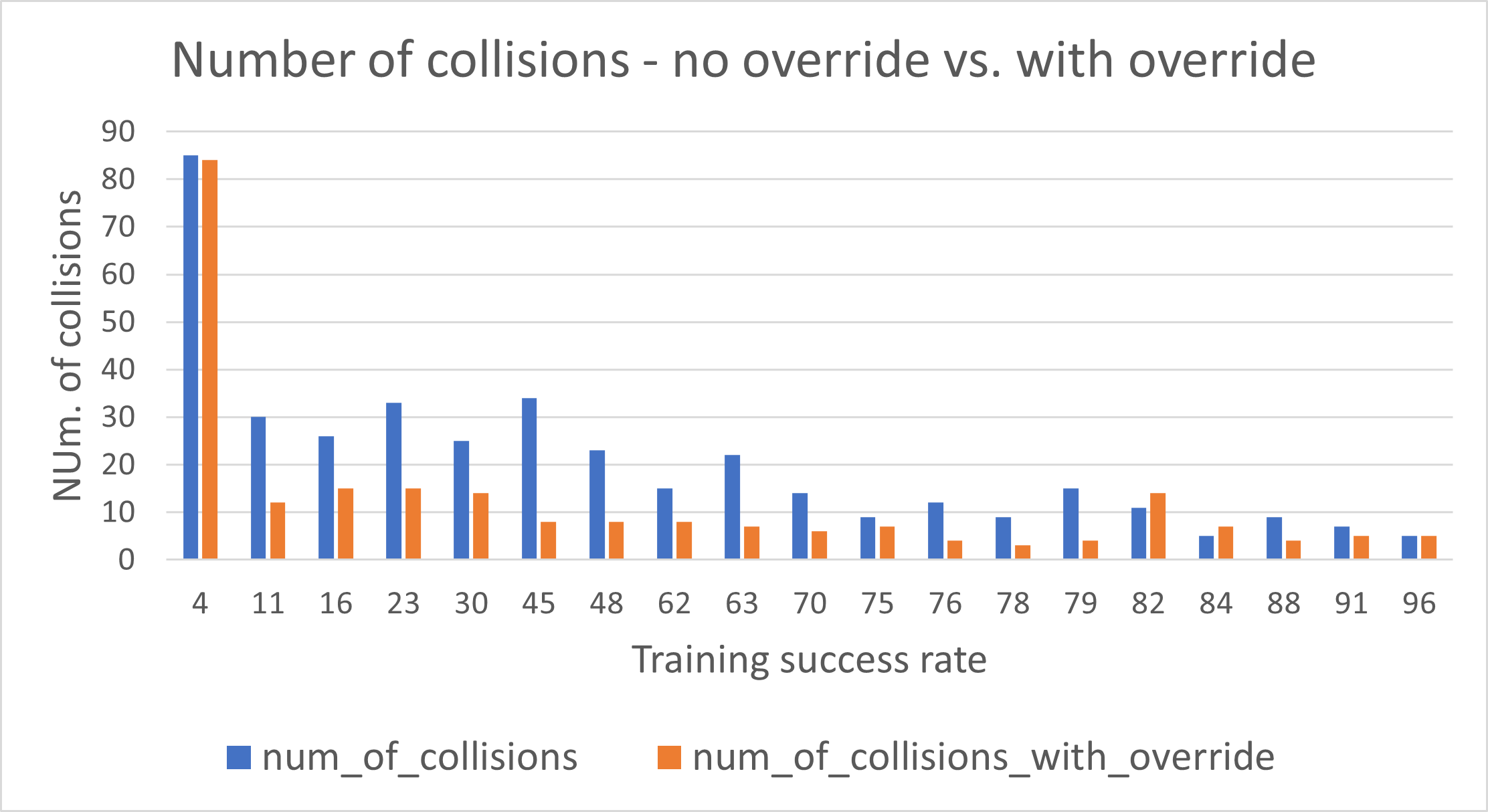}
	\caption{Comparing the number of collisions when the \j{ConservativeAction} override is disabled and then enabled.}
	\label{fig:rob_collisions_compare_override}
\end{figure}

% Describe number of collisions
The experiment's results are summarized in 
Fig.~\ref{fig:rob_collisions_compare_override}, and show that enabling
the override rule leads to a significant reduction in the number of
collisions. We notice that, as the agent's success rate  increases,
the improvement rate decreases. One hypothesis for this behavior could be that ``stronger'' models are more confident in their recommended
actions, thus requiring fewer activations of the override rule.

\begin{figure}[ht]
	\centering
	\includegraphics[width=1.0\linewidth]{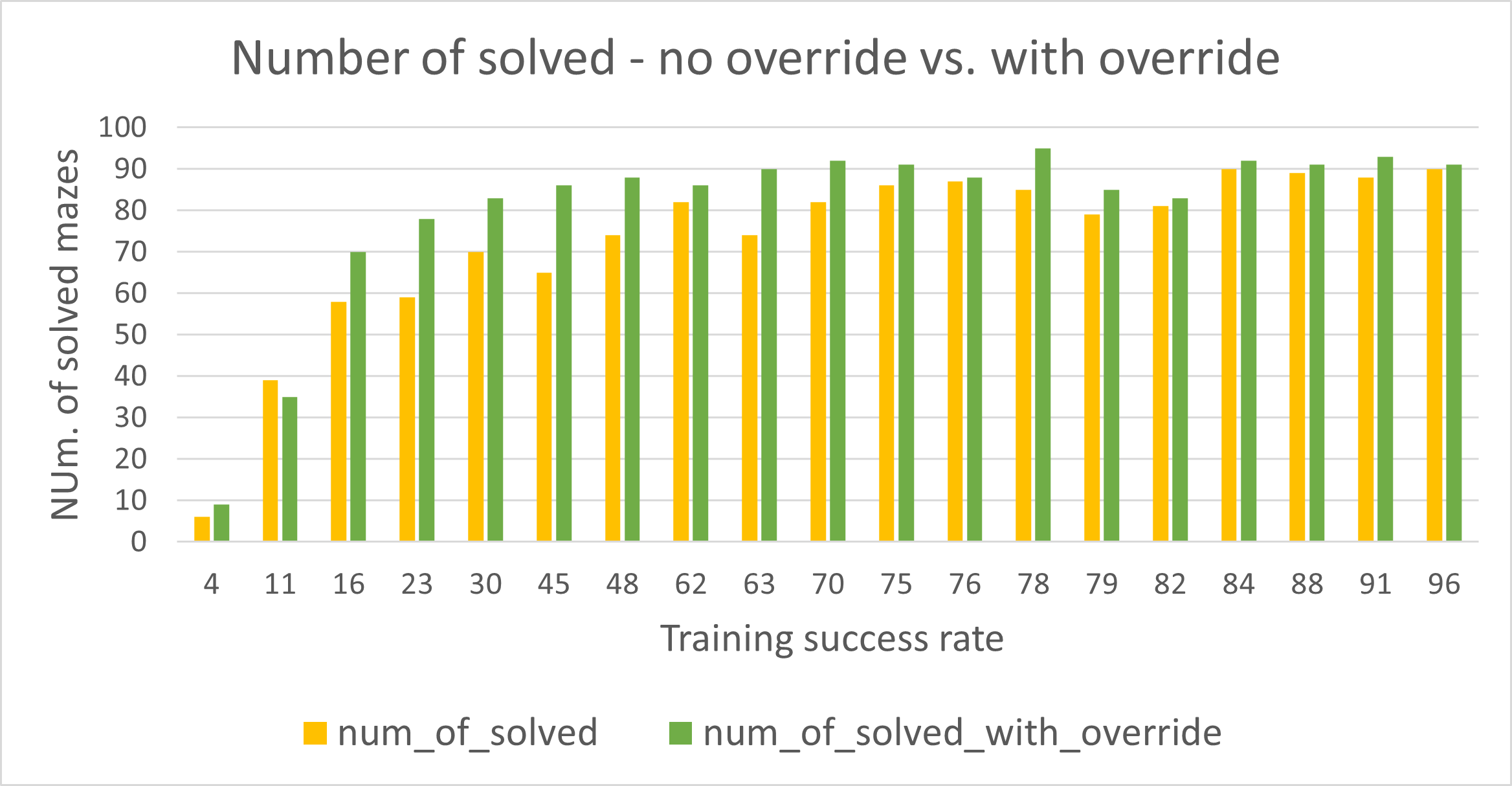}
	\caption{Comparing the num. of solved mazes when the \j{ConservativeAction} override is disabled and then enabled.}
	\label{fig:rob_solved_compare_override}
\end{figure}

% The number of solved mazes
Fig.~\ref{fig:rob_solved_compare_override} portrays a general
improvement in the agent's success rate when the override rule is
enabled, which is (unsurprisingly) correlated with the reduction in
the number of collisions. A possible explanation is that ``mediocre''
agents, i.e.~those with success rates in the range between $16\%$ and
$70\%$, learned policies that are good enough to navigate towards the
target, but which require some assistance in order to avoid obstacles along the 
way.

\begin{figure}[ht]
	\centering
	\includegraphics[width=1.0\linewidth]{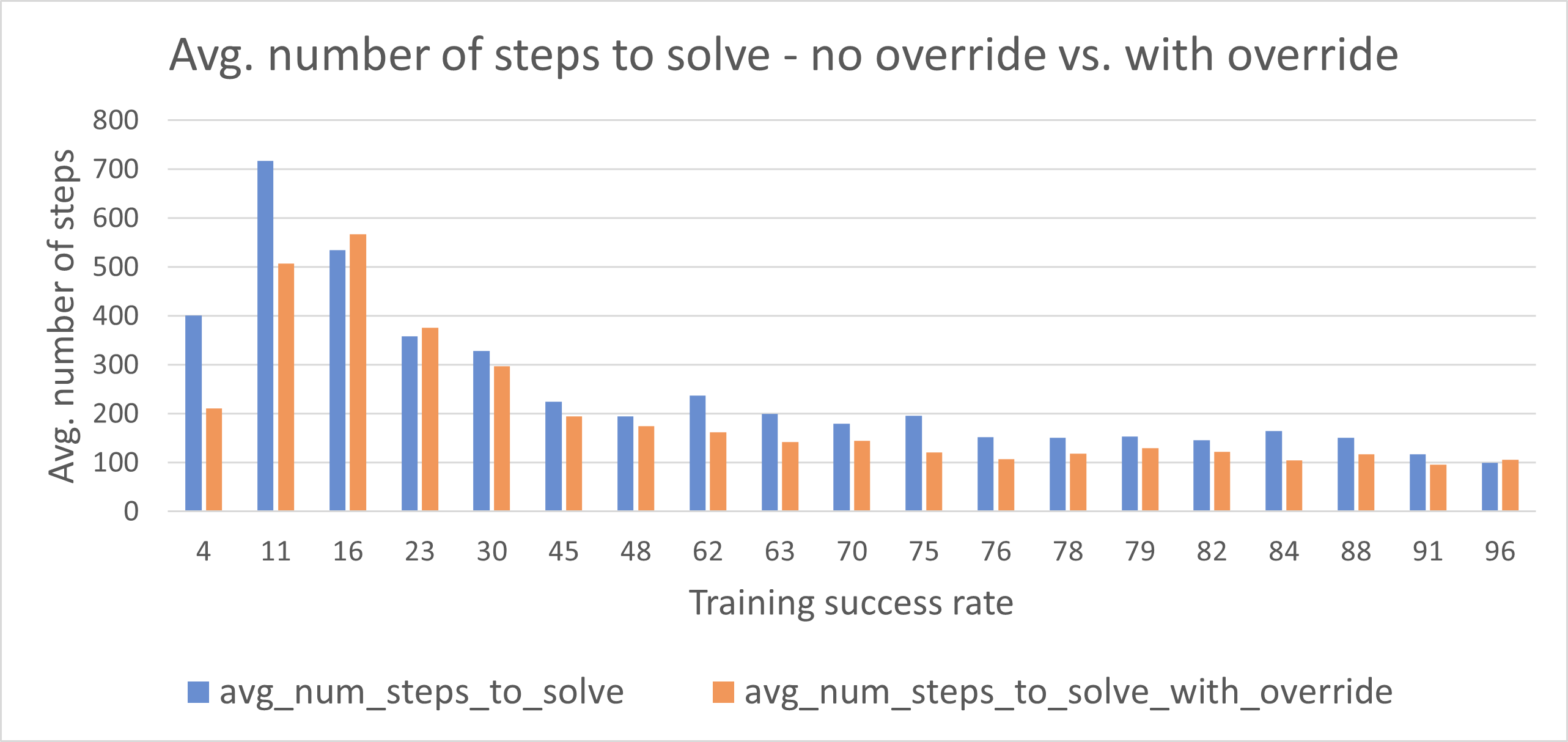}
	\caption{Comparing the average number of steps to solve when the \j{ConservativeAction} override is disabled and then enabled.}
	\label{fig:rob_avg_steps_compare_override}
\end{figure}

% describe avg.number of steps graph
Fig.~\ref{fig:rob_avg_steps_compare_override} depicts a reduction in the
average number of steps required for an agent to solve the maze, when the
override scenario is enabled. This somewhat surprising result
indicates that although our agents can successfully solve mazes, the
\j{ConservativeAction} scenario renders their navigation more
efficient. We speculate that for these agents, selecting actions with
low confidence scores leads to redundant steps.

\section{\uppercase{Introducing Modifier Scenarios}}
\label{sec:modifier-scenario}

\subsection{Motivation}

% The disadvantages of the proxy solution 
In both of our case studies, we needed to create scenarios
capable of reasoning about the events being requested in the current
time step --- which we resolved by introducing new, ``proxy''
events. However, such a solution has several drawbacks. First, it
entails extensive renaming of existing events, and the modification of
existing scenarios, which goes against the incremental nature of
SBM~\cite{HaMaWe12}. Second, once added, the override scenario
becomes a crucial component in the $O_{DNN}$ infrastructure, without
which the system cannot operate; and in the common case where the
override rule is not triggered, this incurs unnecessary
overhead. Third, it is unclear how to support the case where several  scenarios 
are required. These drawbacks indicate that the
``proxy'' solution is complex, costly, and leaves much to be desired.

% What is our desired idiom/construct
In order to address this need and allow users to design override
rules in a more convenient manner, we propose here to extend the
idioms of SBM in a way that will support \emph{modifier scenarios}:
scenarios that are capable of observing and modifying the current
event, as it is being selected for triggering. A formal definition
appears below.

\subsection{Defining Modifier Scenarios}

% Extended SBM definitions
We extend the definitions of SBM from Sec.~\ref{sec:background} with
a new type of scenario, named a \emph{modifier scenario}. A modifier
scenario is formally defined as a tuple
$O_\modifier = \langle Q^{M}, q_0^{M}, \delta^{M}, f^{M}\ra$, where:
\begin{itemize}
\item $Q^M$ is a set of states representing synchronization points.
\item $q_0^{M}$ is the initial state.
\item $\delta^{M}:Q^M \times E \to Q^M$ is a deterministic transition function, indicating how the scenario reacts when an event is triggered.
\item $f^{M}: Q^M \times 2^E \times 2^E \to E$ is a function that maps
  a state, a set of observed requested events, and a set of observed blocked 
  events into an event from the set $E$. $f^{M}$ can operate in a 
  deterministic, well-defined manner, or in a randomized manner to select a 
  suitable event from $E$.
\end{itemize}
Intuitively, the modifier thread can use its function $f^{M}$ at a 
synchronization point to affect the selection of the current event.

Let $M = \{O^1, ..., O^n, O_\modifier\}$ be a behavioral model, where
$n \in N$, each $O^i = \langle Q^i, q_0^i, \delta^i ,R^i, B^i\ra$ is an
ordinary scenario object, and $O_\modifier$ is a modifier scenario object.
In order to define the semantics of $M$, we construct the labeled
transition system $LTS(M) = \langle Q, q_0, \delta, f^{M} \ra$, where:

\begin{itemize}
\item $Q := Q^1 \times ... \times Q^n \times Q^M$ is the set of states.
\item $q_{0} := \langle q_0^1, ..., q_0^n,q_0^M \ra \in Q $ is the initial state.
\item $f^{M}:= f^M$ is the modification function of $O_\modifier$.
\item $\delta: Q \times E \to Q$ is a deterministic transition function, defined for all $q = \langle q^1, ..., q^n,q^M \ra \in Q$ and $e \in E$ by
\begin{equation*}
	\delta(q,e) := 
	\langle\delta^1(q^1,e),...,\delta^n(q^n,e),\delta^M(q^M,e)\ra
\end{equation*}
\end{itemize}

An execution of $P$ is an execution of $LTS(M)$. The
execution starts from the initial state $q_0$, and in each state
$q \in Q$, the event selection mechanism collects the sets of
requested and blocked events, namely
$R(q):=\bigcup\limits_{i=1}^{n} R^i(q^i)$ and
$B(q):=\bigcup\limits_{i=1}^{n} B^i(q^i)$.

The set of enabled events at synchronization point $q$ is $E(q) = R(q) 
\setminus B(q)$. If $E(q) = \emptyset$ then the system is deadlocked. 
Otherwise, the ESM allows the modifier scenario to affect event selection, by 
applying $f^{M}$ and selecting the event:
\[
  e=f^{M}(q, R(q), B(q)).
\]

The ESM then triggers $e$, and notifies the 
relevant scenarios. By convention, we require that $f^{M}$ does not select an 
event that is currently blocked; although it can select events that are not 
currently requested. The state of $LTS(M)$ is then updated according to $e$. 
The execution of $LTS(M)$ is formally recorded as a sequence of triggered 
events (a run). For simplicity, we assume that there is a single
$O_\modifier$ object in the model, although in practice it can be
implemented using a collection of scenarios.

\subsection{Revised Override Scenarios}

We extend the definition of an override rule over a network
$N$, into a tuple
$\langle P, Q, f\rangle$, where:
\begin{inparaenum}[(i)]
\item $P(x)$ is a predicate over the network's input vector $x$;
\item $Q(N(x))$ is a predicate over the network's output vector $N(x)$; and
\item $f:\mathbb{O}\to\mathbb{O}$ is a function that replaces the
  proposed network output event with a new output event.
\end{inparaenum}
Using a modifier scenario  $O_\modifier$, we can now implement this
more general form of an override rule within an SB model. As an illustrative 
example, we change the override rule from
Sec.~\ref{sec:background-overrideRules}, to consider the network's output as
well:
\[
\langle x_1 > x_2, y_2 > 1, f(y_i) \to y_1 \rangle
\]

\begin{figure}[ht]
	\centering
	\includegraphics[scale=0.30]{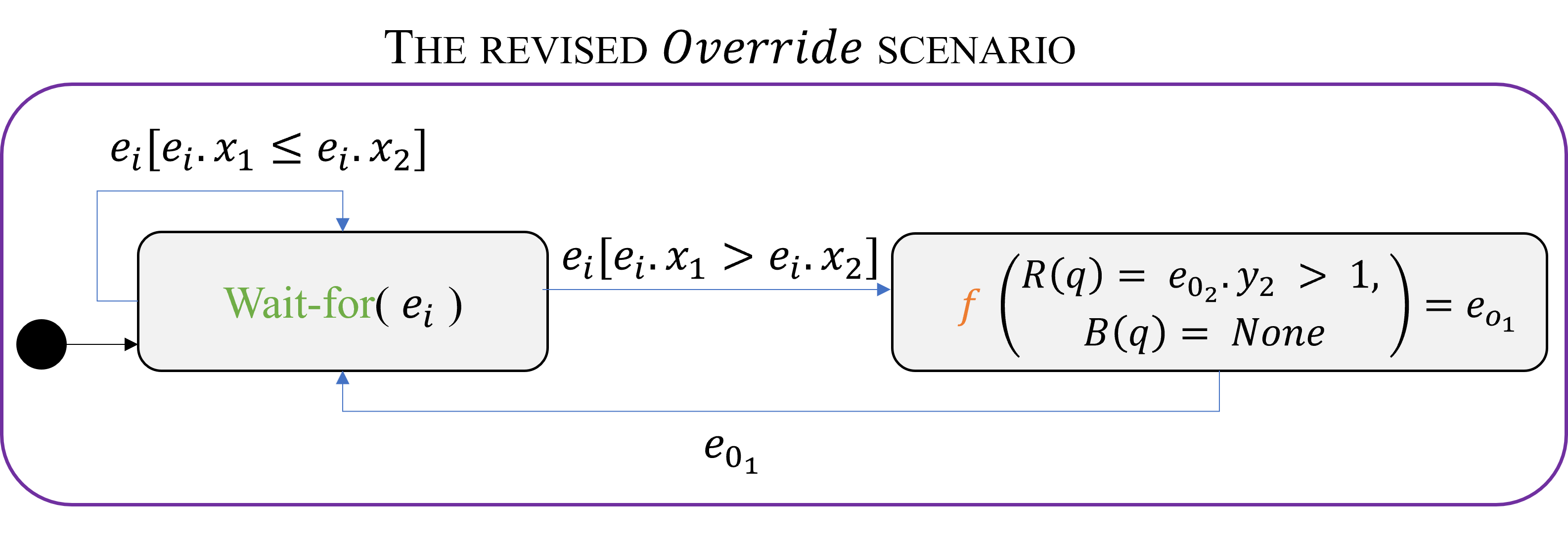} 
	\caption{An $O_\modifier$ scenario object for enforcing the
          override rule that whenever $x_1 > x_2$ and
          $y_2 > 1$, output event $e_{o_1}$ should be triggered. The
          scenario waits for the input event to satisfy the predicate, and then 
          proceeds to the state where it declares a modification. 
          The first argument to the modification function $f$ is the output
          event and assignment that the scenario would like to modify. The 
          second argument to the function is the set of blocked events:
          \emph{None}, in our case. The return value from the function is the 
          output event, $e_{o_1}$. At the synchronization point, the 
          ESM collects the requested and blocked events, applies the $f$ 
          function of the modifier scenario, and then notifies the relevant 
          scenarios that $e_{o_1}$ output event has been selected for 
          triggering.}
	\label{fig:override-rule-with-modify}
\end{figure}

Note that this definition differs from the original: it takes
into account the currently selected output event and its value. Also, it 
contains a function $f$ that, whenever the predicates hold, maps a network-selected output into action  $y_1$. An updated version of the override 
rule, implemented as a modifier scenario, appears in
Fig.~\ref{fig:override-rule-with-modify}. To support the ability to 
observe output event $e_o$'s internal value, the event contains a payload of 
the calculated output neurons' values.

\begin{figure*}[htb]
	\centering
	\includegraphics[width=0.75\linewidth]{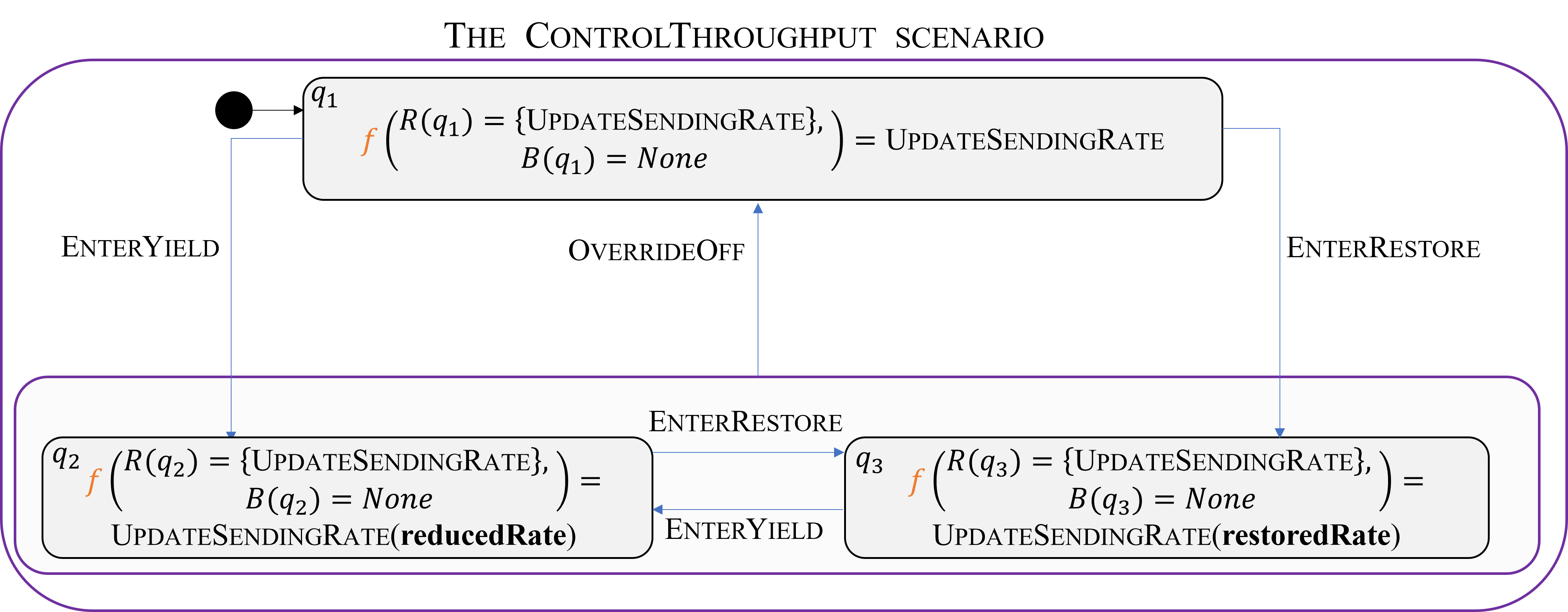}
	\caption{The \j{ControlThroughput} override scenario. The scenario waits-for events 
		\j{\{OverrideOff,EnterYield,EnterRestore\}} in each state, and transitions subsequently. It also observes and possibly modifies the 	
		\j{UpdateSendingRate} event using its $f$ function depending on the 
		current state $q_i$, $R(q_i)$ and $B(q_i)$. E.g., if we are in 
		$q_2$, and \j{UpdateSendingRate} is requested but not blocked, its 
		value will be modified according to the reduce policy. Note that the 
		modification of the \j{UpdateSendingRate} does not result in a 
		transition to a different state.}
	\label{fig:aurora_control_throughput_scenario}
\end{figure*}

\begin{figure*}[htb]
	\centering
	\includegraphics[width=0.75\linewidth]{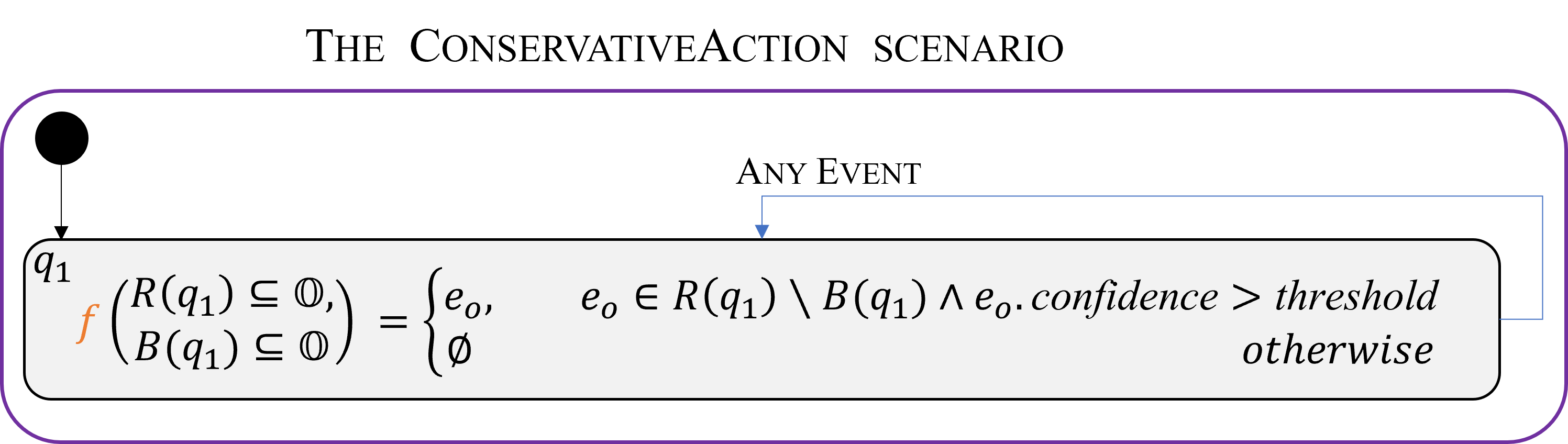}
	\caption{The revised \j{ConservativeAction} scenario. The
		scenario observes all possible subsets 
		of the output events $\mathbb{O}$ that are requested or blocked at
		state $q_{1}$. Concretely, at each synchronization point, the ESM launches 
		$f$ with a specific set of the requested and blocked output events. The 
		function $f$ is only concerned with the set of enabled (requested and 	
		not blocked) output events. From these possible output events, $f$ 	
		randomly selects an event $e_{o}$ according to this policy:
		\begin{inparaenum}[(i)]
			\item  if the selected event is above the threshold, $f$ passes the 
			event as-is; and
			\item if the selected event is below the threshold, $f$ randomly
			selects a different possible output event which is above the 
			threshold.
		\end{inparaenum}
		The ESM then notifies the relevant scenarios of the selected output 
		event. If no such event exists, the program is in a deadlock, in which 
		case the scenario can reduce the threshold to find a possible event. 
		The scenario remains in its state $q_{1}$ whenever any event is 
		triggered.}
	\label{fig:turtlebot_conservative_scenario_v2}
\end{figure*}

% Revising the set of scenarios, and override scenarios, from Aurora.
With the updated override rule definition, we now
refactor the scenarios from Sec.~\ref{sec:scavengerMode}. First, 
we restore 
\j{UpdateSendingRate} to its original role as an output event (as
opposed to a proxy event).
Second, we modify the \j{MonitorNetworkState} scenario to request three 
events that signal the current throughput state:
\begin{inparaenum}[(i)]
	\item \j{OverrideOff}, which signals that the sending rate should be
	forwarded as-is; and
	\item \j{EnterYield} and 
	\item \j{EnterRestore}, which signal that the sending rate should be
	overridden by the yield/restore policy.
\end{inparaenum} Third, we introduce the \j{ControlThroughput} override 
scenario, replacing \j{ReduceThroughput} and \j{RestoreThroughput}. This scenario waits-for a signal on the current throughput 
state, and transitions between the internal states that represent it. 
The scenario uses function $f$ to observe the requested
event \j{UpdateSendingRate} in each state. When the output event 
\j{UpdateSendingRate} is requested, $f$ is executed 
and receives the requested and blocked events as parameters. If the event is 
blocked, we are in a deadlock. If the scenario is in the \j{OverrideOff} state, 
the function returns the event as-is. If the scenario is in the \j{EnterYield/ 
EnterRestore} states, the scenario returns an
\j{UpdateSendingRate} event with a sending rate that is modified according to
the matching policy. The revised \j{UpdateSendingRate} event is then triggered,
and all relevant scenarios proceed with their execution. 
Fig.~\ref{fig:aurora_control_throughput_scenario} depicts the new 
\j{ControlThroughput} scenario.

% Revising the set of scenarios, and override scenarios, from TurtleBot.
We now revise the set of scenarios we implemented to support the 
TRL project~\ref{sec:case_robotis} and the conservative override 
rule~\ref{sec:case_robotis_conservative}. The first modification is to restore 
the \j{OutputEvent} event to its original role as an output, instead
of a proxy event. We then use the $f$ function to simplify the 
\j{ConservativeAction} scenario. Recall that originally, the scenario waited 
for the \j{InputEvent}, for the purpose of re-playing the $O_{DNN}$ evaluation 
if the selected \j{OutputEvent} was below the threshold. The revised scenario 
can define an $f$ that will observe the set of requested output events, and 
then randomly select an output event that exceeds the threshold, and which is 
not blocked. If there are no possible output events, the system is deadlocked. 
From a practical point of view, the scenario can reduce the threshold to avoid 
this situation (assuming that $R(q) \setminus B(q)$ is not empty).
Fig.~\ref{fig:turtlebot_conservative_scenario_v2} displays the revised 
\j{ConservativeAction} scenario.

In summary, we have successfully revised the override rules from our two case 
studies utilizing the $O_\modifier$ extension. First, this new and more 
powerful definition has enabled us to implement the rules without ``proxy'' 
events. This change reduces the high coupling between the scenarios of the 
original implementation. Second, the redesigned models offer a more compact and 
direct approach:
\begin{inparaenum}[(i)]
	\item the two override rules from the Aurora case study were
          reduced to a single scenario; and
	\item the TRL override rule contains a single synchronization point. 
\end{inparaenum}
These characteristics are more in line with the SBM spirit that views  
scenarios as simple and self-contained components. Moving forward, we plan to
enhance the existing  SBM packages with the $O_\modifier$ extension.
	
\section{\uppercase{Related Work}}
\label{sec:related-work}
\noindent
 
Override rules are becoming an integral part of many DRL-based
systems~\cite{Ka20a}. The concept is closely related to that
of \emph{shields} and \emph{runtime monitors}, which have been
extensively used in the field of robotics~\cite{PhYaGrSmSt17},
drones~\cite{DeGhSeShTi18}, and many
others~\cite{HaMoSc06,FaMoFeRi11,ScDeRiGaCoHoStSm15,JiLa17,WuRaRaLaSe18}. We
regard our work as another step towards the goal of  effectively creating, and
maintaining, override rules for complex systems.

Although our focus here has been on designing override rules using
SBM,  other modeling
formalisms could be used just as well.  Notable examples include
the publish-subscribe framework~\cite{EuFeGuKe03}, aspect oriented
programming~\cite{KiLaMeMaLoLoIr97}, and the BIP
formalism~\cite{BlSi08}. A key property of SBM, which seems to render
it a good fit for override rules, is the native idiom support for
blocking events~\cite{Ka20a}; although similar support could
be obtained, using various constructs, in other formalisms.

\section{\uppercase{Conclusion}}
\label{sec:conclusion}
\noindent

As DNNs are increasingly being integrated into complex systems, there
is a need to maintain, extend and adjust them --- which has given rise
to the creation of override rules. In this work, we
sought to contribute to the ongoing effort of facilitating the
creation of such rules, through two extensive case studies. Our
efforts exposed a difficulty in an existing, SBM-based method for
designing guard rules, which we were then able to mitigate by
extending the SBM framework itself. We hope that this effort, and
others, will give rise to formalisms that are highly equipped for
supporting engineers in designing override rules for DNN-based systems.

\section*{\uppercase{Acknowledgements}}

We thank the anonymous reviewers for their insightful comments. This work was 
partially supported by the Israeli Smart Transportation Research Center (ISTRC).

\bibliographystyle{apalike}
{\small
	\bibliography{references}
}

\end{document}